%% file: main.tex
\documentclass[reprint,superscriptaddress,amsmath,amssymb,amsart,aps]{revtex4-1}

\usepackage{graphicx}
\usepackage{dcolumn}
\usepackage{bm}
\usepackage{hyperref}
\usepackage{xcolor}
\usepackage{tikz}

\usepackage{xr}
\usepackage{cleveref}
\usepackage{multibib}
\newcites{SM}{SM References}
\Crefname{figure}{Fig.}{Figs.}
\Crefname{equation}{Eq.}{Eqs.}

\definecolor{lime}{HTML}{A6CE39}
\DeclareRobustCommand{\orcidicon}{
	\begin{tikzpicture}
	\draw[lime, fill=lime] (0,0) 
	circle [radius=0.16] 
	node[white] {{\fontfamily{qag}\selectfont \tiny ID}};
	\draw[white, fill=white] (-0.0625,0.095) 
	circle [radius=0.007];
	\end{tikzpicture}
	\hspace{-2mm}
}
\foreach \x in {A, ..., Z}{\expandafter\xdef\csname orcid\x\endcsname{\noexpand\href{https://orcid.org/\csname orcidauthor\x\endcsname}
			{\noexpand\orcidicon}}
}

\newcolumntype{C}[1]{>{\centering\arraybackslash}p{#1}}

\usepackage{hyperref}

\begin{document}
\include{letter}


\pagebreak
\clearpage
\widetext

\include{supplemental}

\end{document}

%% file: letter.tex
\title{Flavor Dependence of Charged Pion Fragmentation Functions}
\newcommand*{\MSU }{Mississippi State University, Mississippi State, Mississippi 39762, USA}
\newcommand*{\MSUindex}{1}
\affiliation{\MSU}
\newcommand*{\WM }{The College of William \& Mary, Williamsburg, Virginia 23185, USA}
\newcommand*{\WMindex}{2}
\affiliation{\WM}
\newcommand*{\TEMP }{Temple University, Philadelphia, Pennsylvania 19122, USA}
\newcommand*{\TEMPindex}{3}
\affiliation{\TEMP}
\newcommand*{\ANL }{Argonne National Laboratory, Lemont, Illinois 60439, USA}
\newcommand*{\ANLindex}{4}
\affiliation{\ANL}
\newcommand*{\JLAB }{Thomas Jefferson National Accelerator Facility, Newport News, Virginia 23606, USA}
\newcommand*{\JLABindex}{5}
\affiliation{\JLAB}
\newcommand*{\BOULDER }{University of Colorado Boulder, Boulder, Colorado 80309, USA}
\newcommand*{\BOULDERindex}{6}
\affiliation{\BOULDER}
\newcommand*{\YER }{A.I. Alikhanyan  National  Science  Laboratory \\ (Yerevan  Physics
Institute),  Yerevan  0036,  Armenia}
\newcommand*{\YERindex}{7}
\affiliation{\YER}
\newcommand*{\CUA }{Catholic University of America, Washington, DC 20064, USA}
\newcommand*{\CUAindex}{8}
\affiliation{\CUA}
\newcommand*{\REG }{University of Regina, Regina, Saskatchewan S4S 0A2, Canada}
\newcommand*{\REGindex}{9}
\affiliation{\REG}
\newcommand*{\ZAG }{University of Zagreb, Zagreb, Croatia}
\newcommand*{\ZAGindex}{10}
\affiliation{\ZAG}
\newcommand*{\HU }{Hampton University, Hampton, Virginia 23669, USA}
\newcommand*{\HUindex}{11}
\affiliation{\HU}
\newcommand*{\CNU }{Christopher Newport University, Newport News, Virginia 23606, USA}
\newcommand*{\CNUindex}{12}
\affiliation{\CNU}
\newcommand*{\UVA }{University of Virginia, Charlottesville, Virginia 22903, USA}
\newcommand*{\UVAindex}{13}
\affiliation{\UVA}
\newcommand*{\NCAT }{North Carolina A \& T State University, Greensboro, North Carolina 27411, USA}
\newcommand*{\NCATindex}{14}
\affiliation{\NCAT}
\newcommand*{\SUNO }{Southern University at New Orleans, New Orleans, Louisiana 70126, USA}
\newcommand*{\SUNOindex}{15}
\affiliation{\SUNO}
\newcommand*{\UTENN }{University of Tennessee, Knoxville, Tennessee 37996, USA}
\newcommand*{\UTENNindex}{16}
\affiliation{\UTENN}
\newcommand*{\UCONN }{University of Connecticut, Storrs, Connecticut 06269, USA}
\newcommand*{\UCONNindex}{17}
\affiliation{\UCONN}
\newcommand*{\ODU }{Old Dominion University, Norfolk, Virginia 23529, USA}
\newcommand*{\ODUindex}{18}
\affiliation{\ODU}
\newcommand*{\OHIO }{Ohio University, Athens, Ohio 45701, USA}
\newcommand*{\OHIOindex}{19}
\affiliation{\OHIO}
\newcommand*{\UOY }{University of York, Heslington, York, Y010 5DD, UK}
\newcommand*{\UOYindex}{20}
\affiliation{\UOY}
\newcommand*{\FIU }{Florida International University, University Park, Florida 33199, USA}
\newcommand*{\FIUindex}{21}
\affiliation{\FIU}
\newcommand*{\JMU }{James Madison University, Harrisonburg, Virginia 22807, USA}
\newcommand*{\JMUindex}{22}
\affiliation{\JMU}
\newcommand*{\VM}{Virginia Military Institute, Lexington, Virginia 24450, USA}
\newcommand*{\VMindex}{23}
\affiliation{\VM}
\newcommand*{\SBU }{Stony Brook University, Stony Brook, New York 11794, USA}\newcommand*{\SBUindex}{24}
\affiliation{\SBU}
\newcommand*{\TU}{Tsinghua University, Beijing 100084, China}\newcommand*{\TUindex}{25}
\affiliation{\TU}

\newcommand{\footremember}[2]{%
   \footnote{#2}
    \newcounter{#1}
    \setcounter{#1}{\value{footnote}}%
}
\newcommand{\footrecall}[1]{%
    \footnotemark[\value{#1}]%
}

\author{H.\,Bhatt}\affiliation{\MSU}
\author{P.\,Bosted}\affiliation{\WM}
\author{S.\,Jia}\affiliation{\TEMP}
\author{W.\,Armstrong}\affiliation{\ANL} 
\author{D.\,Dutta}\affiliation{\MSU}
\author{R.\,Ent}\affiliation{\JLAB}   
\author{D.\,Gaskell}\affiliation{\JLAB} 
\author{E.\,Kinney}\affiliation{\BOULDER} 
\author{H.\,Mkrtchyan}\affiliation{\YER}

\author{S.\,Ali}\affiliation{\CUA}
\author{R.\,Ambrose}\affiliation{\REG} 
\author{D.\,Androic}\affiliation{\ZAG}  
\author{C.\,Ayerbe Gayoso}\affiliation{\MSU} 
\author{A.\,Bandari}\affiliation{\WM}   
\author{V.\,Berdnikov}\affiliation{\CUA}    
\author{D.\,Bhetuwal}\affiliation{\MSU}
\author{D.\,Biswas}\affiliation{\HU}  
\author{M.\, Boer}\affiliation{\TEMP}
\author{E.\,Brash}\affiliation{\CNU}    
\author{A.\,Camsonne}\affiliation{\JLAB}
\author{M.\,Cardona}\affiliation{\TEMP}          
\author{J.\,P.\,Chen}\affiliation{\JLAB}           
\author{J.\,Chen}\affiliation{\WM}    
\author{M.\,Chen}\affiliation{\UVA}             
\author{E.\,M.\,Christy}\affiliation{\HU}          
\author{S.\,Covrig}\affiliation{\JLAB}           
\author{S.\,Danagoulian}\affiliation{\NCAT}     
\author{M.\,Diefenthaler}\affiliation{\JLAB}     
\author{B.\,Duran}\affiliation{\TEMP}            
\author{M.\,Elaasar}\affiliation{\SUNO}
\author{C.\,Elliot}\affiliation{\UTENN}
\author{H.\,Fenker}\affiliation{\JLAB}           
\author{E.\,Fuchey}\affiliation{\UCONN}           
\author{J.\,O.\,Hansen}\affiliation{\JLAB}           
\author{F.\,Hauenstein}\affiliation{\ODU}       
\author{T.\,Horn}\affiliation{\CUA}             
\author{G.\,M.\,Huber}\affiliation{\REG}       
\author{M.\,K.\,Jones}\affiliation{\JLAB}          
\author{M.\,L.\,Kabir}\affiliation{\MSU}
\author{A.\,Karki}\affiliation{\MSU}            
\author{B.\,Karki}\affiliation{\OHIO} 
\author{S.\,J.\,D.\,Kay}\affiliation{\REG}\affiliation{\UOY}
\author{C.\,Keppel}\affiliation{\JLAB}           
\author{V.\,Kumar}\affiliation{\REG}
\author{N.\,Lashley-Colthirst}\affiliation{\HU}        
\author{W.\,B.\,Li}\affiliation{\WM}               
\author{D.\,Mack}\affiliation{\JLAB}              
\author{S.\,Malace}\affiliation{\JLAB}           
\author{P.\,Markowitz}\affiliation{\FIU}       
\author{M.\,McCaughan}\affiliation{\JLAB}
\author{E.\,McClellan}\affiliation{\JLAB}
\author{D.\,Meekins}\affiliation{\JLAB}          
\author{R.\,Michaels}\affiliation{\JLAB}         
\author{A.\,Mkrtchyan}\affiliation{\YER}        
\author{G.\,Niculescu}\affiliation{\JMU}        
\author{I.\,Niculescu}\affiliation{\JMU}        
\author{B.\,Pandey}\affiliation{\HU}\affiliation{\VM}           
\author{S.\,Park}\affiliation{\SBU}             
\author{E.\,Pooser}\affiliation{\JLAB}           
\author{B.\,Sawatzky}\affiliation{\JLAB}          
\author{G.\,R.\,Smith}\affiliation{\JLAB}             
\author{H.\,Szumila-Vance}\affiliation{\JLAB}\affiliation{\FIU}
\author{A.\,S.\,Tadepalli}\affiliation{\JLAB}
\author{V.\,Tadevosyan}\affiliation{\YER}        
\author{R.\,Trotta}\affiliation{\CUA}           
\author{H.\,Voskanyan}\affiliation{\YER}
\author{S.\,A.\,Wood}\affiliation{\JLAB}            
\author{Z.\, Ye}\affiliation{\ANL}\affiliation{\TU}
\author{C.\,Yero} \affiliation{\FIU}  
\author{X.\,Zheng}\affiliation{\UVA}        
\collaboration{for the Hall C SIDIS Collaboration}
\noaffiliation

\date{\today}

\begin{abstract}

We have measured the flavor dependence of multiplicities for $\pi^+$ and $\pi^-$ production in semi-inclusive deep-inelastic scattering (SIDIS) on proton and deuteron targets to explore a possible charge symmetry violation in fragmentation functions. The experiment used an electron beam with energies of 10.2 and 10.6 GeV at Jefferson Lab and 
the Hall-C spectrometers. 
 The electron kinematics spanned the range $0.3<x<0.6$, $2<Q^2<5.5$ GeV$^2$, and $4<W^2<11$ GeV$^2$. The pion fractional momentum range was 0.3$< z <$0.7, and the transverse momentum range was $0<p_T<0.25$~GeV/c. Assuming factorization at low $p_T$ and allowing for isospin breaking, we find that the results can be described by two ``favored" and two ``un-favored" effective low $p_T$ fragmentation functions that are flavor-dependent. However, they converge to a common flavor-independent value at the lowest $x$ or highest $W$ of this experiment.   
\end{abstract}

      
\maketitle

Semi-inclusive deep-inelastic lepton-nucleon scattering ($l N \rightarrow l^{'} h X$) is an excellent tool to study the quark hadronization mechanism described by fragmentation functions (FF)~\cite{FF1}.
These FF describe how the quarks and gluons transform into color-neutral hadrons or photons during high-energy (hard) scattering processes. Pion semi-inclusive deep-inelastic scattering (SIDIS) is one such scattering process that allows access to the FF associated with the pions identified in the final state. As FF are intrinsically linked to confinement in quantum chromodynamics (QCD), studies of FF are critical for a complete understanding of the basic properties of QCD. Further, FF are the non-perturbative ingredient of the QCD factorization theorems~\cite{factor1} used to analyze hard scattering processes and thereby provide insight into fundamental soft QCD quantities~\cite{FF2}.
The current knowledge of pion FF is based on global QCD analyses~\cite{hkns, DSS1, DSS2, AKK08, JAM, MAP22} that are dominated by measurements from inclusive electron-positron ($e^{+}e^{-}$) annihilation into charged pions at very high energy scales (center-of-mass energy $>$ 10 GeV). Inclusive $e^{+}e^{-}$ annihilation is a clean process to study FF since it is independent of the parton distribution functions (PDF). However, it cannot distinguish between the light quark flavors or the quark and anti-quark FF. Thus, it cannot provide information about possible flavor dependence of FF -- essential for a complete picture of FF as well as the spin structure of nucleons, in particular the transverse spin structure~\cite{signori2013}. 
One of the most important advantages of SIDIS is the ability to constrain the flavor of the quark involved in the scattering process. Consequently, measuring the SIDIS process on protons and deuterons allows an independent extraction of the flavor dependence of FF.
The SIDIS experiments conducted over the last decade have convincingly established that the collinear picture of the quark-parton model is too simple, highlighting the importance of the transverse structure of the hadrons. The flavor structure of FF is important to understand the flavor dependence of the transverse-momentum-dependent (TMD) FF~\cite{signori2013}, and the relative differences between the observed single spin asymmetries of pions and kaons~\cite{hermes09,compass15}. 
Thus, SIDIS measurements provide a unique capability to study the flavor structure of FF at an energy scale that is complementary to that of $e^{+}e^{-}$ annihilation.

It is challenging to model FF as they are non-perturbative objects that cannot be deduced from first principles. 
Current models treat hadronization either as sequential emission of hadrons from colored partons with emission probability parameterized to describe experimental data, such as the Lund string model~\cite{lund}, or approximate it as the emission of a single hadron and an on-shell spectator quark~\cite{Muld96}. Another recent approach uses a combination of these two methods by calculating the emission probability within a QCD-inspired spectator model instead of a parameterization~\cite{NJL-jet}. 
As charge conjugation symmetry (CC) and charge/isospin symmetry (CS/IS) are fundamental properties of QCD and strong interaction processes, 
most models use a simple quark flavor-independent (for light quarks) and isospin-independent ansatz. At the quark level, CS refers to the up ($u$) and down ($d$) quark interactions being identical when their mass difference is neglected~\cite{miller90}. It arises from the invariance of the QCD Hamiltonian under rotations about the 2-axis in isospin space, i.e. 
the interchange of $u$ and $d$ quarks while simultaneously interchanging protons and neutrons~\cite{londergan98}. 
Therefore, CC and CS/IS allow one to drastically reduce the number of independent FF for the light quarks from eight to two~\cite{FFrev}.
As the fragmentation process is a dominantly strong interaction process, FF are expected to respect CS/IS to high precision. 
Most global fits of existing data that extract FF either assume CS or find no significant violation of CS~\cite{DSS1, DSS2}. On the other hand, the transverse polarization of the $\Lambda$ hyperon in $e^+e^-$ annihilation, as measured by the Belle Collaboration~\cite{Belle} seem to indicate a significant IS violation in the corresponding FF. Further, a recent global analysis has reported a significant flavor dependence of FF~\cite{pandm}, posing a significant challenge to QCD. 
These results, 
 and the quest for TMD FF has created an urgent need for a systematic study of the flavor dependence of FF and their charge (isospin) symmetry violation (CSV). 
 SIDIS is well suited for such studies, as the sum and difference ratio of $\pi^+$ and 
$\pi^-$ production on hydrogen to those produced from deuterium serves as an effective test of CS/IS.
In order 
to exploit these advantages, a new SIDIS experimental program was undertaken at the upgraded JLab~\cite{propocsv,propoptsidis,clas12}. An integral part of this program, featuring measurements on both hydrogen and deuterium targets over a wide range of kinematics~\cite{propocsv,propoptsidis}, was completed in 2019. In this letter, we report the results of the tests of charge and isospin symmetry violation and flavor dependence of the unpolarized FF extracted from the SIDIS experimental program. Any flavor dependence of FF would also be significant for other parts of the SIDIS program, such as the test of CSV in PDF~\cite{propocsv}.\\    
\begin{figure*}[!htb]
\includegraphics[width=0.9\textwidth]{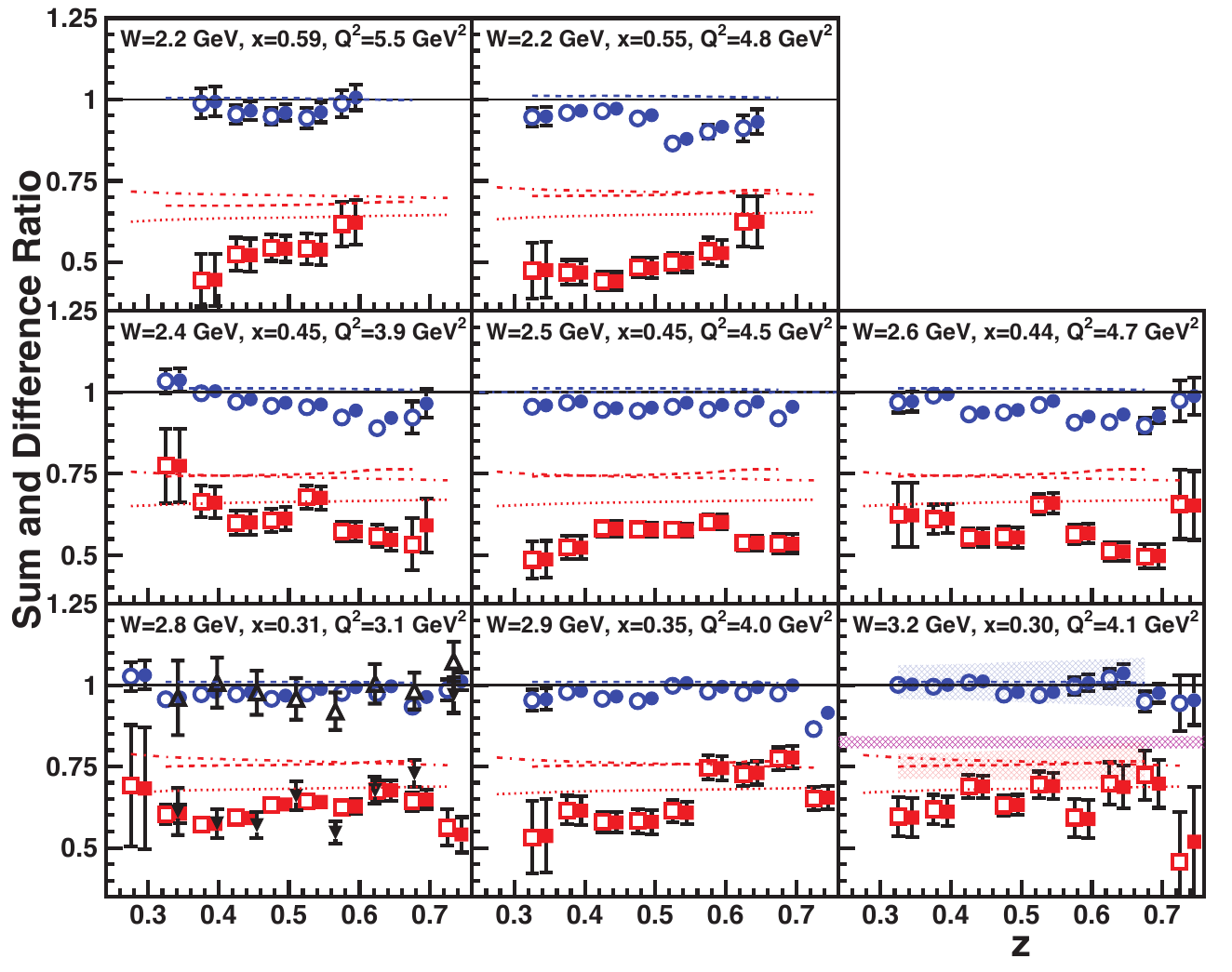}
    \caption{The sum ratio $R_1(z)$ (circles) and difference ratio $R_2(z)$ (squares) as a function of $z$. The eight panels are ordered in increasing values of $W$ when going from left to right with similar values of $x$ for each row. The open (closed) symbols are with (without) subtraction of the diffractive $\rho^0$ contributions. The black solid (red dotted) lines are the predictions for any model with charge/isospin symmetry for the sum (difference) ratio. 
    The dashed curves use FF from the MAP~\cite{lhapdf} collaboration integrated over the same $p_T$ range as the experiment and the dot-dashed curves are FF from the DSS~\cite{DSS1,DSS2} fits integrated over all $p_T$. The blue (red) hatched bands in the bottom-right panel show the
    uncertainty of the MAP curves. 
    The open (closed) triangles in the bottom-left panel show the sum (difference) ratio obtained from the previous JLab 6 GeV experiment~\cite{Tigran07}. 
    The magenta cross hatched band in the bottom-right panel shows the 2.2\% systematic uncertainty of these ratios.}
    \label{fig:sumdiff}
\end{figure*}
The $p_T$-integrated ($p_T$ is the pion transverse momentum) semi-inclusive pion electroproduction yield ($\frac{dN}{dz}$) as a function of the pion's longitudinal momentum fraction, $z$, is usually modeled as 
\begin{equation}
\label{eq:yield}
    \frac{dN}{dz} \sim \sum_i e_{i}^2 q_{i}(x,Q^2)D_{q_i \rightarrow \pi}(z,Q^2), 
\end{equation}
 where the quarks of flavor $i$ with charge $e_i$ carrying a fraction $x$ of nucleon momentum are represented by the PDF, $q_i(x,Q^2)$, 
and the spin averaged FF by $D_{q_i \rightarrow \pi}(z,Q^2)$. 
 As a consequence of collinear factorization~\cite{factor1}, the PDF are independent of $z$ and FF are independent of $x$, but depend on the virtuality scale, or 4-momentum transferred squared ($Q^{2}$), via a logarithmic evolution~\cite{factor1,Tigran07}.

 We define the measured multiplicities for $\pi^{+}$ and $\pi^{-}$ production from hydrogen (p) and deuterium (d) targets, $M^{\pi^{\pm}}_{\mbox{p/d}}(x,Q^2,z)$, as the ratio of the respective SIDIS cross section to the DIS cross section (see Eq.~\ref{eq:bprm} in the online Supplementary Material~\cite{supl}).
 At leading order, 
 assuming CS, and no difference in the $p_T$ dependence of the measured multiplicities, we can write two simple ratios, 
\begin{equation}
R_1(z) = \frac{M_{\mbox{d}}^{\pi^{+}}(z)  + M_{\mbox{d}}^{\pi^{-}}(z)}{M_{\mbox{p}}^{\pi^{+}}(z)  + M_{\mbox{p}}^{\pi^{-}}(z)} = 1
\label{eq:sumratio}
\end{equation}
and
\begin{equation}
R_2(z) = \frac{M_{\mbox{d}}^{\pi^{+}}(z)  - M_{\mbox{d}}^{\pi^{-}}(z)}{M_{\mbox{p}}^{\pi^{+}}(z)  - M_{\mbox{p}}^{\pi^{-}}(z)} 
= \frac{3\left(4u(x)+d(x)\right)}{5\left(4 u(x) - d(x)\right)},
\label{eq:diffratio}
\end{equation}
where 
the $u(d)$ quark PDF are written as
$u(x) = u_{v}(x) + \bar{u}(x) {{\mbox{~~and~~}}} d(x) = d_{v}(x) + \bar{d}(x), $
 with $u_v(d_v)$ and $\bar{u}(\bar{d})$ as the valence quark and sea anti-quark contributions, respectively. Here the quark and anti-quark contributions from the sea are assumed symmetric and the strange quark contributions are neglected. For measurements made in the valence region ($ x >$ 0.3) where the contributions from the sea quarks can be neglected, 
 both ratios are independent of $z$ and $p_T$. Thereby, these two ratios constitute an excellent test of CS  within the collinear factorization formalism~\cite{factor1}.

Most global analyses to extract PDF assume IS and CS in the PDF, which reduces the number of independent PDF by half. If we assume CS in the PDF but allow for non-zero CSV in FF, the multiplicity $M^{\pi^{\pm}}_{\mbox{p/d}}(x,Q^2,z)$
for each target (H/D) and charged pion type can be written in terms of two favored FF, $D_{u\pi^{+}}(z)$, $D_{d\pi^{-}}(z)$, and two un-favored FF, $D_{d\pi^{+}}(z)$, 
 $D_{u\pi^{-}}(z)$, respectively 
 (see Eq.~\ref{eq:s1} in the Supplementary Material~\cite{supl}). Any difference between the two favored and the two un-favored FF is an indication of CSV in FF. The degree of CSV in the favored and un-favored FF can be quantified in terms of two parameters defined as: 
\begin{equation}
\label{eq:ff_asym}
\delta_{\mbox{CSV}}^{f}(z) = \frac{D_{d\pi^{-}}-D_{u\pi^{+}}}{D_{u\pi^{+}}}, ~~ \delta_{\mbox{CSV}}^{uf}(z) = \frac{D_{d\pi^{+}}-D_{u\pi^{-}}}{D_{u\pi^{-}}}
\end{equation}
Most current global analyses to extract FF either impose exact CS, or arrive at CSV parameters which are effectively zero. 

We have measured the four $p_T$-integrated multiplicities for the electroproduction of $\pi^{\pm}$ from hydrogen and deuterium targets. These multiplicities, along with the PDF from a global fit of world data
 were used to extract the four FF. We have assumed an identical $p_T$ dependence for the $\pi^{\pm}$ multiplicities from hydrogen and deuterium, integrated over $p_T$ with an average of $<p_T> ~=~ $~0.1~GeV/c.
 The CSV of FF are quantified in terms of the two parameters in Eq.~\ref{eq:ff_asym}. 

 \begin{figure}[!hbt]
\includegraphics[width=0.45\textwidth]{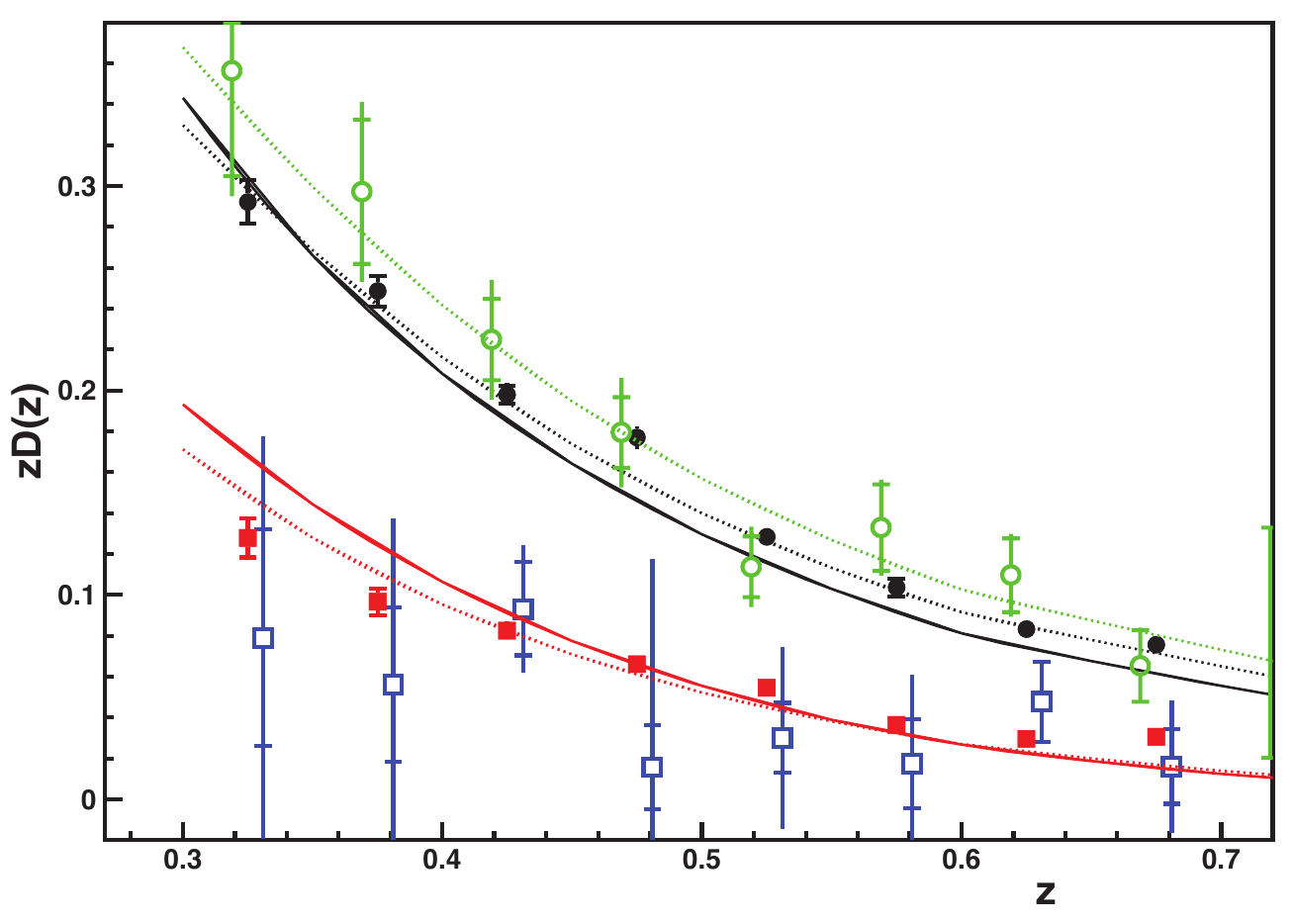}
\caption{The $z$ dependence of the two favored FF ($D_{u \pi^{+}}$ black solid and $D_{d \pi^{-}}$ green open, circles) and two un-favored FF ($D_{u \pi^{-}}$ red solid and $D_{d \pi^{+}}$ blue open, squares) extracted from the multiplicities without assuming CS/IS. The results for the highest $W$ (lowest $x$) setting ($W$=3.2 GeV, $x$= 0.3) are shown. The inner error bars are the statistical uncertainty while the outer error bars are the total uncertainty which includes the systematic uncertainty. The solid lines are FF from the JAM collaboration~\cite{JAM}, while the dashed lines are FF from the DSS collaboration~\cite{DSS1,DSS2}. The open points have been shifted in $z$ for clarity.
}
\label{fig:ff_kin8}
\end{figure}

 The experiment was carried out in the Fall of 2018 and the Spring of 2019, in Hall C at JLab. The experiment used the quasi-continuous wave electron beam with beam 
energies of 10.2 and 10.6~GeV and beam currents ranging from 2 $\mu$A to 70~$\mu$A.  Additional details of the experiment 
are described in Sec.~\ref{sec:supsec1} of the Supplementary Material~\cite{supl}. 
The experimental yields were obtained from selected electron-pion coincidence events per milli-Coulomb of electrons incident on $^1$H, and $^2$H targets.  The selected events passed cuts on momentum, scattering angles, and missing mass of the residual system, $M_X$, where $M_X$ was restricted to be above the resonance region ($M_X >$ 1.6 $\mathrm{GeV}/\mathrm{c}^2$).
The yields were integrated over azimuthal angle ($\phi$) and $p_T$. 
 The backgrounds from the target's aluminum windows and accidental coincidences were subtracted. This normalized SIDIS pion electroproduction yield was corrected for all known inefficiencies of the two spectrometers such as the detector efficiencies (97\%–99\%), trigger efficiency (98\%-99\%), tracking efficiencies, computer and electronic live times (94\%–99\%). The
corrected yields were binned in $z$ for the 8 different kinematic settings where the $x$ ranged from 0.3 to 0.6, $Q^2$ ranged from 3.1 to 5.5 GeV$^2$ and the center-of-mass energy, $W$, ranged from 2.2 to 3.2 GeV. The table of kinematic settings is shown in Table ~\ref{tab:kinematics} of the Supplementary Material~\cite{supl}.

\begin{figure}[!hbt]
\includegraphics[width=0.45\textwidth]{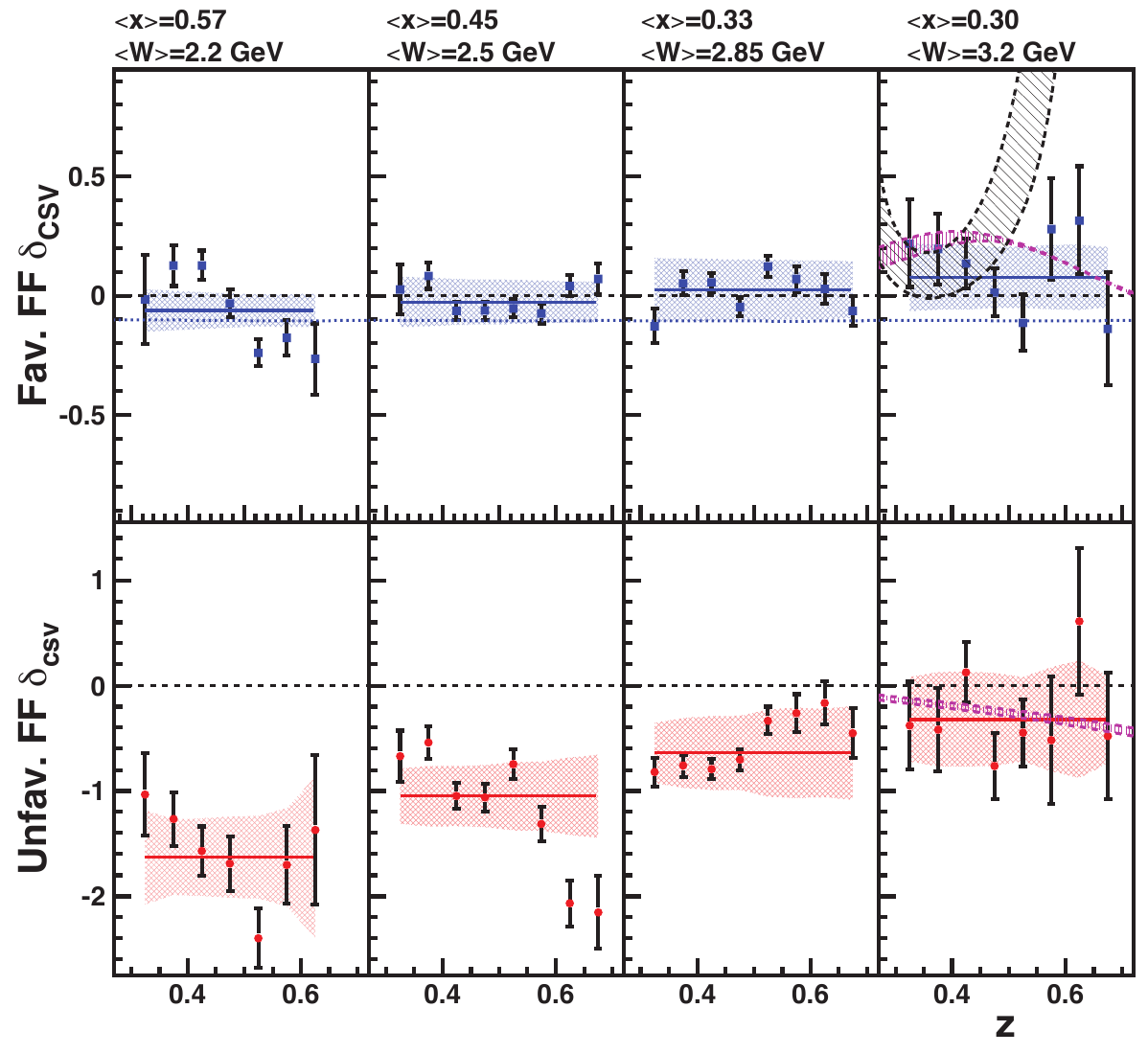}
\caption{The $z$ dependence of the charge/isospin symmetry violating parameter $\delta_{CSV}$ for the favored FF (top panels) and un-favored FF (bottom panels), extracted from the measured charged pion multiplicities on hydrogen and deuterium targets. Horizontally, the panels are ordered in decreasing values of $x$ (increasing $W$). The blue (red) solid lines are a constant value fit to the favored (un-favored) $\delta_{CSV}$. The shaded bands are the systematic uncertainty. 
Assuming charge symmetry, the $\delta_{CSV}$ should be zero, as indicated by the black dashed lines. The magenta band with vertical hatching in the right-most panels is the $\delta_{CSV}$ and its uncertainty from fits by Peng and Ma~\cite{pandm}, the black band with angled hatching is the $\delta_{CSV}$ and its uncertainty predicted by the MAP collaboration~\cite{lhapdf},
while the blue dotted lines (top panels only) are from the DSS collaboration~\cite{DSS1,DSS2}.}
\label{fig:ff_asym}
\end{figure}

A Monte Carlo (MC) simulation~\cite{simc} of the SIDIS process was performed with the factorized form shown in Eq.~\ref{eq:yield}. The CTEQ5 next-to-leading-order (NLO) PDF were used to parametrize $q(x,Q^2)$~\cite{cteq} along with a parametrization of FF from global fits of SIDIS data~\cite{bosted_fit}.  
The MC was used to smear parameterized PDF and FF over the experimental acceptance. The MC included corrections due to radiative tails from exclusive pion electroproduction, pion decay, and electroproduction of $\rho^0$ mesons and $\Delta$(1232) resonances. Additional details about the models used in the simulation can be found in Refs.~\cite{hem_thesis,shuo_thesis,ptsidis_pub}. 
The Monte Carlo yields were integrated over the same phase space as the measured yields.

The corrected experimental yields for the 8 different settings 
were used along with the Monte Carlo yield and the model cross section to obtain the four multiplicities,  $M^{\pi^{\pm}}_{\mbox{p/d}}(x,Q^2,z)$. The systematic uncertainty of the extracted multiplicities is 2.8\%, it is listed in Table~\ref{tab:syst} and described in the Supplementary Material~\cite{supl}.
The four multiplicities as a function of $W^2$ are shown in Fig.~\ref{fig:4mults} of the Supplementary Material~\cite{supl} for 8 different $z$ bins ranging from $z = $ 0.3 to 0.7. Note that our data confirm that the $p_T$ dependence for the $\pi^{\pm}$ multiplicities from hydrogen and deuterium are identical within the small $p_T$ range covered.
The four multiplicities were used to form the two ratios $R_1(z)$ and $R_2(z)$, and were also used to obtain four FF ($D_{u\pi^{+}}(z), D_{d\pi^{-}}(z)$ and $D_{u\pi^{-}}(z), D_{d\pi^{+}}(z)$) by simultaneously solving a system of four 
 equations (Eq.~\ref{eq:s1} in the Supplementary Material~\cite{supl}).

The extracted sum and difference ratios and their statistical uncertainties are shown as a function of $z$ in 
Fig.~\ref{fig:sumdiff}. For the sum and difference ratios, many systematic uncertainties cancel to first order resulting in a net 2.2\% systematic uncertainty shown by the magenta band in the bottom-right panel (see the Supplemental Material~\cite{supl} for additional details).
The open (closed) symbols are with (without) subtraction of the diffractive $\rho^0$ contributions. The negligible difference between them shows that the diffractive $\rho^0$ contribution to the pion yield has very little impact on these ratios. The solid (dotted) lines are the expectations for models with CS/IS. The dot-dashed and dashed curves use FF from the global fits by the DSS~\cite{DSS1, DSS2} and MAP~\cite{lhapdf} collaborations, respectively. The uncertainty for the DSS curves is not shown because, unlike the experiment and MAP results, they are integrated over all $p_T$. At the highest $W$ (3.2 GeV) the 
two ratios are remarkably independent of $z$ over the entire range ($z$ = 0.3 - 0.7) and are also consistent with the magnitude predicted by the global fits 
to existing data. In other words, the results agree with the CS/IS expectation. The sum ratio $R_1$ slowly but steadily deviates from the CS expectation with decreasing~$W$ (increasing~$x$), both in terms of 
the $z$ independence and the magnitude. Similarly, the difference ratio also 
shows increasingly large deviations from the CS expectation with decreasing $W$. These deviations may indicate the importance 
of higher twist contributions to the SIDIS cross sections at low $W$. These results also indicate that even for the limited range of $p_T$ 
covered in this experiment, CS/IS seems to be valid for $W >$ 3 GeV.
Moreover, the sum/difference ratio from the previous JLab 6 GeV 
experiment~\cite{Tigran07} (shown as black triangles in the bottom left panel) agrees remarkably well with the current results. These 
older ratios were obtained at the same $x$ = 0.32, but at significantly lower 
$W$ and $Q^2$ of 2.4 GeV and 2.3 GeV$^2$ respectively. This seems to indicate 
that $x$ may also be relevant for tests of CS/IS.

The four multiplicities were also used to extract four FF which are shown as a function of $z$ in Fig.~\ref{fig:ff_kin8} for the highest $W$ (lowest $x$) setting ($W$= 3.2 GeV, $x$=0.3). Note that the JAM collaboration (solid lines) assumes CS/IS for all FF, while the DSS collaboration (dashed lines) assumes CS/IS only for the un-favored FF. 
The variation of the extracted FF due to the normalization type uncertainties of the multiplicities was used to determine the systematic uncertainty of the FF. The extracted FF for all 8 settings are shown in Fig.~\ref{fig:4ffs} of the Supplementary Material~\cite{supl}.
The two favored and two un-favored FF were used to form the favored and un-favored $\delta_{CSV}(z)$ parameters as defined in Eq.~\ref{eq:ff_asym} and are shown in Fig.~\ref{fig:ff_asym}. The settings with similar $W$ and $x$ were averaged to reduce the 8 kinematic settings to 4. The variation in the $\delta_{CSV}$ parameter due to the choice of PDF and normalization type uncertainty was used to determine the systematic uncertainties of $\delta_{CSV}$. The systematic uncertainty is shown by the shaded bands.  
The favored $\delta_{CSV}$ parameter is essentially zero within the experimental uncertainties over the entire range of $z$ and $W$. 
They are also consistent with the expectations of the global fits by DSS~\cite{DSS2} and Peng and Ma~\cite{pandm} but not with the fits by the MAP collaboration~\cite{lhapdf}.

The statistical uncertainties of the un-favored $\delta_{CSV}$ parameter is significantly 
larger than those for the favored. Within these large uncertainties, the un-favored $\delta_{CSV}$ is consistent with zero at the highest 
 $W$ but deviates from zero with  decreasing~$W$ (increasing~$x$). These results and the sum and difference 
ratios shown in Fig.~\ref{fig:sumdiff} are a direct experimental confirmation of CS/IS for both the favored and un-favored FF at the highest $W$. 
The results confirm that for $W >$ 3 GeV ($x \leq $ 0.35), the FF are flavor-independent, 
and the fragmentation process obeys CS/IS within experimental uncertainties. The results also show a more complex fragmentation process at lower $W$ (higher $x$), with possible contributions from higher-order corrections. 

The poor 
statistics in the un-favored down quark fragmentation channel drive the larger uncertainty in the un-favored CSV parameter. Even in an 
isoscalar target, up quark scattering is a majority of the DIS cross section
due to a larger electromagnetic coupling, and the poor statistics are exacerbated for 
SIDIS by the un-favored fragmentation configuration. 
Lacking a free neutron target, tagging the spectator (A-1) system would isolate hard 
scattering on the neutron. High-luminosity measurements with the spectator 
tagging of a proton or $^3$He (using a D or $^4$He target respectively) could significantly improve the uncertainties for un-favored down quark fragmentation.

In summary, we have measured the $\pi^{\pm}$ multiplicities from SIDIS on 
hydrogen and deuterium targets over a wide range of kinematics. The sum and difference ratios of the four multiplicities satisfy  CS/IS at the highest $W$ (3.2 GeV) but steadily deviate from the CS expectation with decreasing~$W$ (increasing~$x$). The 
multiplicities were also used to quantify the flavor dependence of FF, they confirm the flavor independence of both the favored and un-favored FF at the highest $W$.
The favored FF are flavor independent over the $W$ range of the experiment. The un-favored FF have an increasing flavor dependence with decreasing~$W$, albeit with large experimental uncertainty. The results also
indicate that higher-twist corrections are important for low $W$. When the data reported here are included in future global fits of PDF and FF including higher-order corrections, they will provide further detailed insight into the fragmentation process.
These results also suggest that CSV in FF is less likely to interfere with the forthcoming extraction of CSV in PDF~\cite{propocsv} from 
the deuteron data at high $W$ (low $x$).
The spectator tagging technique pioneered at JLab could be used in future experiments to access nearly free neutron targets to improve the precision of the unfavored FF and their CSV.

This work was funded in part by the U.S. Department of Energy, including contract 
AC05-06OR23177 under which Jefferson Science Associates, LLC operates Thomas Jefferson National Accelerator Facility, and by the U.S. Department of Energy, Office of Science, contract numbers DE-AC02-06CH11357, DE-FG02-07ER41528, DE-FG02-96ER41003, and by the U.S. National Science Foundation grants PHY 2309976, 2012430 and 1714133 and the Natural Sciences and Engineering Research Council of Canada grant SAPIN-2021-00026. We wish to thank the staff of Jefferson Lab for their vital support throughout the experiment. We are also grateful to all granting agencies providing funding support to authors throughout this project.

\bibliographystyle{apsrev4-1}
\bibliography{main}

%% file: supplemental.tex
\setcounter{equation}{0}
\setcounter{figure}{0}
\setcounter{table}{0}
\setcounter{section}{0}
\setcounter{page}{1}
\makeatletter
\renewcommand{\theequation}{S\arabic{equation}}
\renewcommand{\thefigure}{S\arabic{figure}}
\renewcommand{\thetable}{S\arabic{table}}
\renewcommand{\thesection}{S-\Roman{section}}
\renewcommand{\bibnumfmt}[1]{[S#1]}
\renewcommand{\citenumfont}[1]{S#1}

\begin{center}
\textbf{\large Supplementary Material for Flavor Dependence of Charged Pion Fragmentation Functions}
\end{center}

\section{The Experiment}
\label{sec:supsec1}
The experiment was carried out in Hall C at Jefferson Lab using a quasi-continuous wave electron beam with energies of 10.2 to 10.6~GeV and beam currents ranging from 2 $\mu$A to 70~$\mu$A.
The beam energy was measured with $<0.05\%$ relative uncertainty from the bend angle of the beam as it traversed a set of magnets with precisely known field integrals.
The total accumulated beam charge was determined using a set of resonant-cavity based beam-current monitors and a parametric transformer as gain monitor. 
The relative uncertainty of the accumulated beam charge was $\approx$ 0.5\%, after correcting for zero-offsets and saturation effects measured using beam current scans on a solid carbon target. 
The beam was rastered at $\approx$ 25 kHz over a 2$\times$2~mm$^2$ square pattern to minimize density reduction in the target due to localized beam heating. 

The main production targets were a 10-cm-long (726 mg/cm$^2$) liquid hydrogen and a 10-cm-long (1690 mg/cm$^2$) liquid deuterium targets. 
Two aluminum foils placed 10-cm apart were used to determine the background from the 
aluminum entrance ($\approx$ 14 mg/cm$^2$) and exit ($\approx$ 19 mg/cm$^2$) end caps of the cryogenic target cells. 
A small reduction in density due to localized beam heating was determined to be -0.023\%/$\mu A$ for the liquid hydrogen target and -0.027\%/$\mu A$ for the liquid deuterium target.
\begin{table*}[!hbt]
\caption{The eight kinematic settings where data were collected on both hydrogen and deuterium targets.} 
\label{tab:kinematics}\centering 
 \begin{tabular}{cccccccc}
\hline\hline
Ebeam & $E^{'}$& $\theta_{e}$ &$Q^{2}$ & $W$ & $x$ & $p_{\pi}$ & $\theta_{\pi}$ \\
\hline
($GeV$) & ($GeV/c$) & (deg) &($GeV^{2}$)  & ($GeV$)&  & ($GeV/c$) & ($deg$)\\
\hline 
10.2 &	5.240 & 18.51 &5.5 & 2.2 & 	0.59 & 2.219, 2.713, 3.208 & 17.75\\
10.6 &  5.971 & 15.75 &4.8 & 2.2 & 0.55 & 1.838, 2.299, 2.761, 3.223  & 18.55 \\
10.6 &	5.971 &	14.24 &3.9 & 2.4 & 0.45 & 1.838, 2.299, 2.761, 3.223 & 17.04\\
10.6 &	5.240 &	16.30 &4.5 & 2.5 & 0.45 & 2.525, 3.363, 5.04 & 8-26\\
10.6 &	4.945 &	17.26 & 4.7 & 2.6 & 0.44 & 2.241, 2.804, 3.366, 3.928  & 14.16\\
10.6 &	5.240 &	13.50 &3.1 & 2.8 & 0.31 & 1.956, 2.575, 3.433, 4.79 & 8-30\\
10.6 &	4.483 &	16.64 &4.0 & 2.9 & 0.35 & 2.428, 3.037, 3.646, 4.234 & 11.61\\
10.6 &	3.307 &	19.70 &4.1 & 3.2 & 0.30 &  2.645, 3.393, 4.531, 6.786& 8-22\\
\hline 
\hline
\end{tabular}
\end{table*}  

Scattered  electrons were detected in the High Momentum Spectrometer~\cite{s1_cal_ref} in coincidence with charged pions detected in the Super High Momentum 
Spectrometer~\cite{s2_shms_nim}.  
The angle and momentum of the electron arm (13 - 49 deg., 1 - 6 GeV/c) and the hadron arm (6 - 30 deg., 2 - 7 GeV/c) were chosen to map the region between 0.2~$\leq x \leq$~0.6 and 0.3~$\leq z \leq$~0.7, where $x$ is the fraction of nucleon momentum carried by the struck quark, and $z$ is the pion's longitudinal momentum fraction. The angle, $\theta_{pq}$, between the 
electron three-momentum transfer, $\vec{q}$ and the hadron momentum, 
was chosen to cover a range in pion transverse momentum $p_{T}$ up to 0.25 GeV/c. The kinematics 
of the experiment are listed in Table~\ref{tab:kinematics}.\\

\begin{figure*}[bth!]
\includegraphics[width=0.45\textwidth]{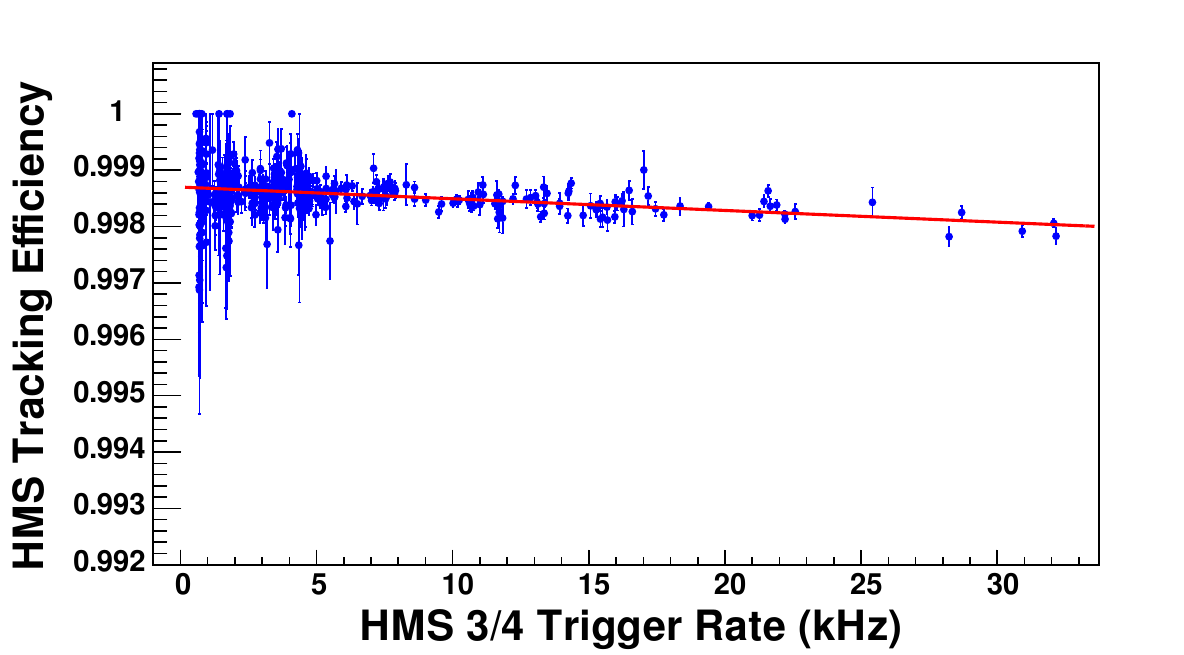}
\includegraphics[width=0.45\textwidth]{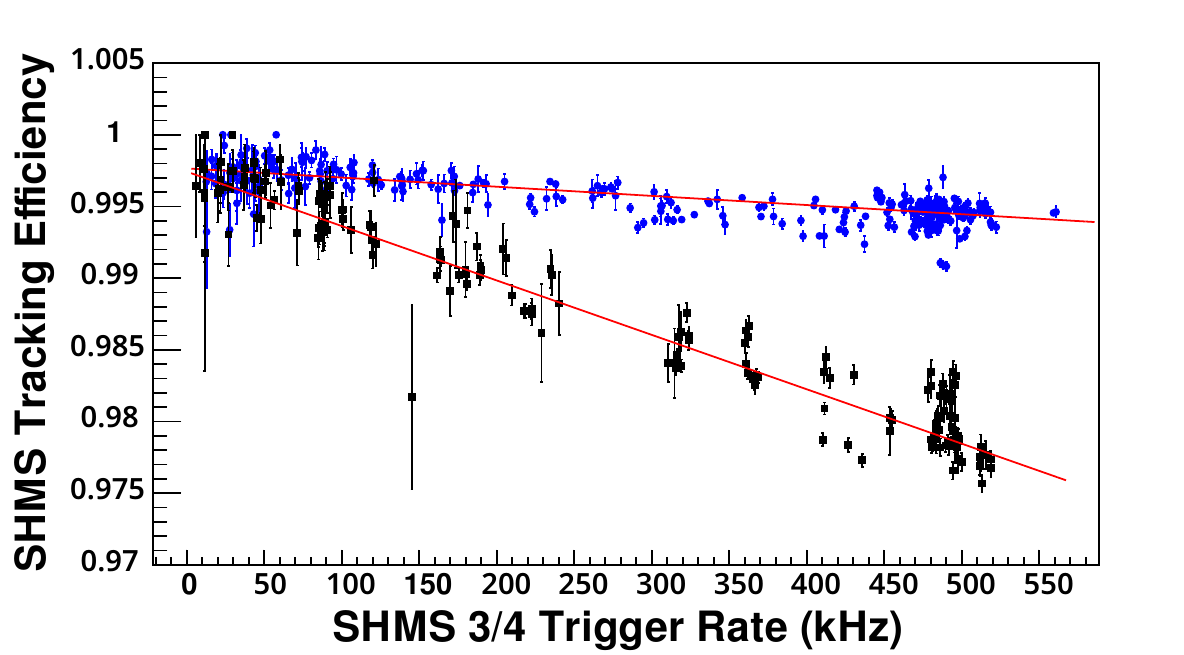}
\caption{ The tracking efficiency of the HMS (left) and SHMS (right) drift chambers 
as a function of the 3/4 trigger rate. The rate dependence of the efficiency is fit to a first order polynomial. For the HMS, the $\chi^2$ per degree of freedom is 1.2. For the SHMS, the $\chi^2$ per degree of freedom is 7.9
for the Spring 2018 (black squared) and 1.2 for the Fall 2019 (blue circles) run periods.
The tracking efficiency corrections were applied run-by-run and 
only the statistical uncertainties are shown.}
\label{fig:hms_shms_tracking}
\end{figure*}

The detector packages of the two spectrometers are similar, and they 
included four segmented planes of plastic scintillators (except for the last 
plane in the SHMS which used quartz bars) that were used to form 
the trigger in order to read out the time and amplitude signals from all of the detectors. 
To ensure nearly 100\% efficiency for the triggers, signals from any three out of the four planes in each spectrometer were required. Henceforth referred to as the 3/4 trigger for each spectrometer. The time resolution of each plane was about
0.5 nsec, resulting in an accuracy of typically 0.3 nsec when all four planes
were combined. 
Two drift chambers, each containing six planes of wires oriented at 
0$^{\circ}$ and $\pm$60$^{\circ}$ with respect to the horizontal, 
provided position and direction (track) information at the 
spectrometer focal plane with a resolution of $<$250~$\mu$m. 
The track information was used to reconstruct the momentum 
and the angle of the particle at the target (reaction vertex). After many improvements to the tracking software, the tracking  
efficiency in the HMS was determined to be over 99.7\% throughout the experiment as shown in Fig.~\ref{fig:hms_shms_tracking} (left).
For the SHMS, the tracking efficiency varied between 99.5\% at low trigger rates to 98\% at the highest trigger rate. The rate dependence of the tracking efficiency was slightly different between the Spring 2018 and Fall 2019 run periods, as shown in Fig.~\ref{fig:hms_shms_tracking} (right).

\begin{figure*}[hbt!]
\includegraphics[width=0.45\textwidth]{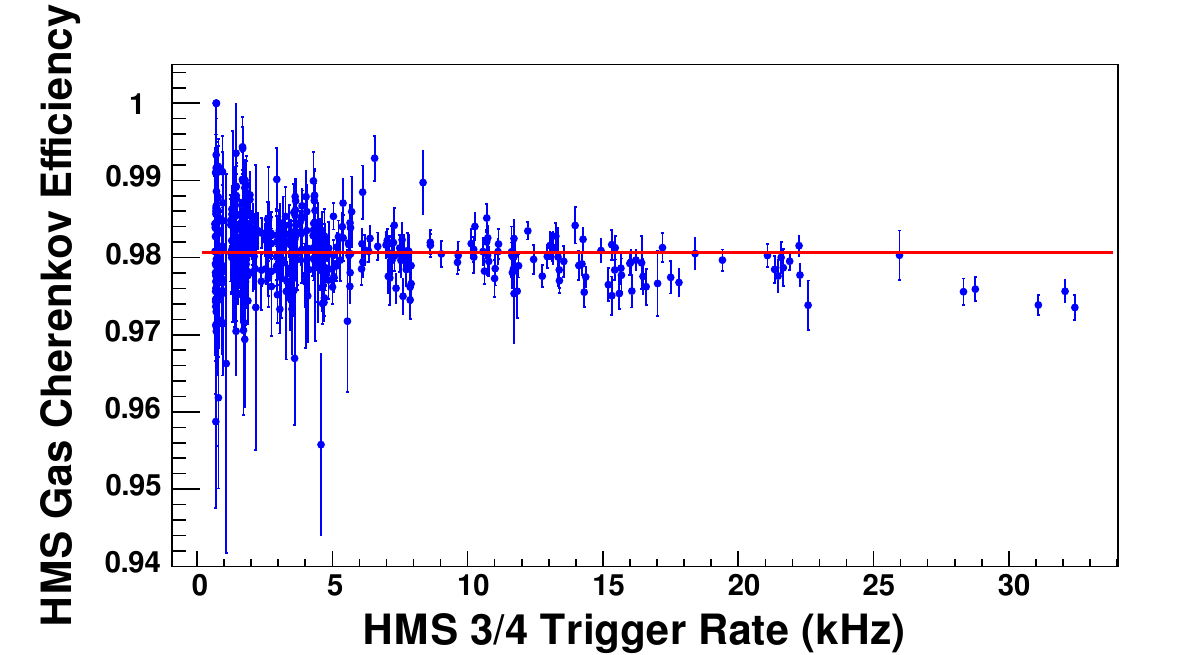}
\includegraphics[width=0.45\textwidth]{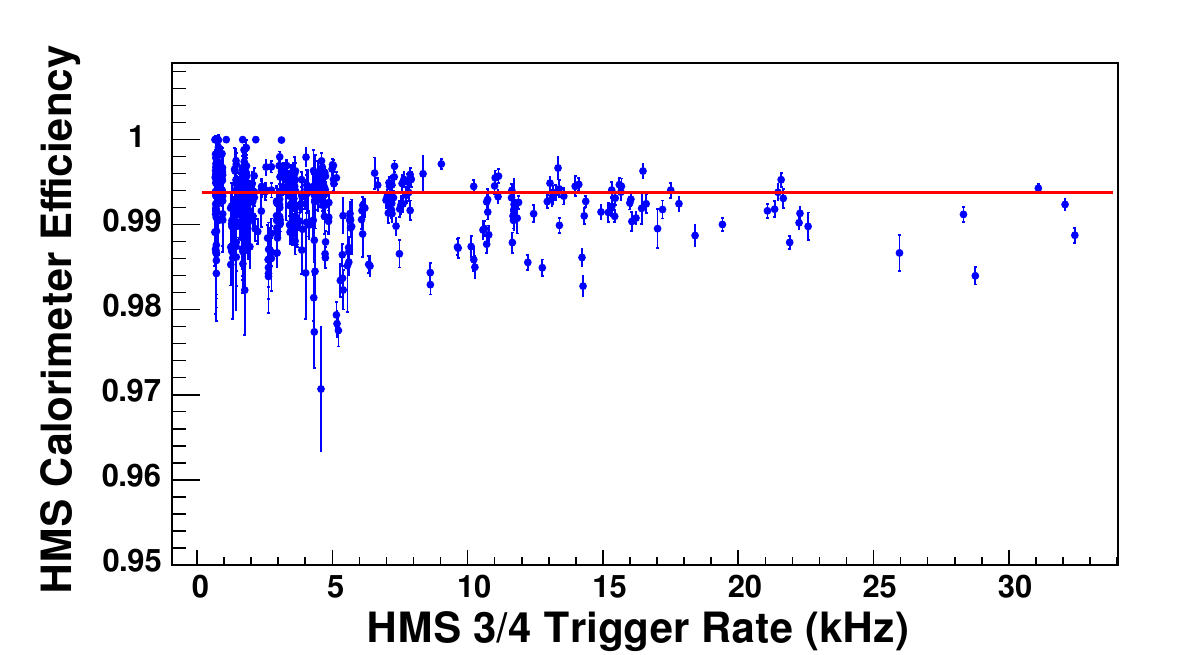}
\caption{ The HMS gas Cherenkov efficiency (left) and the HMS calorimeter efficiency (right) as a function of HMS 3/4 trigger rate. 
The solid lines show the constant value fits for each, with a $\chi^2$ per degree-of-freedom of 1.7 and 9.9, respectively.
For the HMS gas Cherenkov, a constant value of 0.98 was used as the correction factor, while a constant value of 0.994 was used for the calorimeter. Only the statistical uncertainties are shown.}
\label{fig:hms_cer_cal_eff}
\end{figure*}

In the HMS (the electron spectrometer), a threshold gas Cherenkov detector and a 
segmented Pb-glass calorimeter~\cite{s1_cal_ref} were used for electron 
identification. A constant efficiency of 98\% was estimated for the Cherenkov detector in the HMS, as shown in Fig.~\ref{fig:hms_cer_cal_eff} (left). The efficiency of the HMS calorimeter was $\sim$99\% throughout the experiment as shown in Fig.~\ref{fig:hms_cer_cal_eff} (right). 

\begin{figure*}[htb!]
\includegraphics[width=0.9\textwidth]{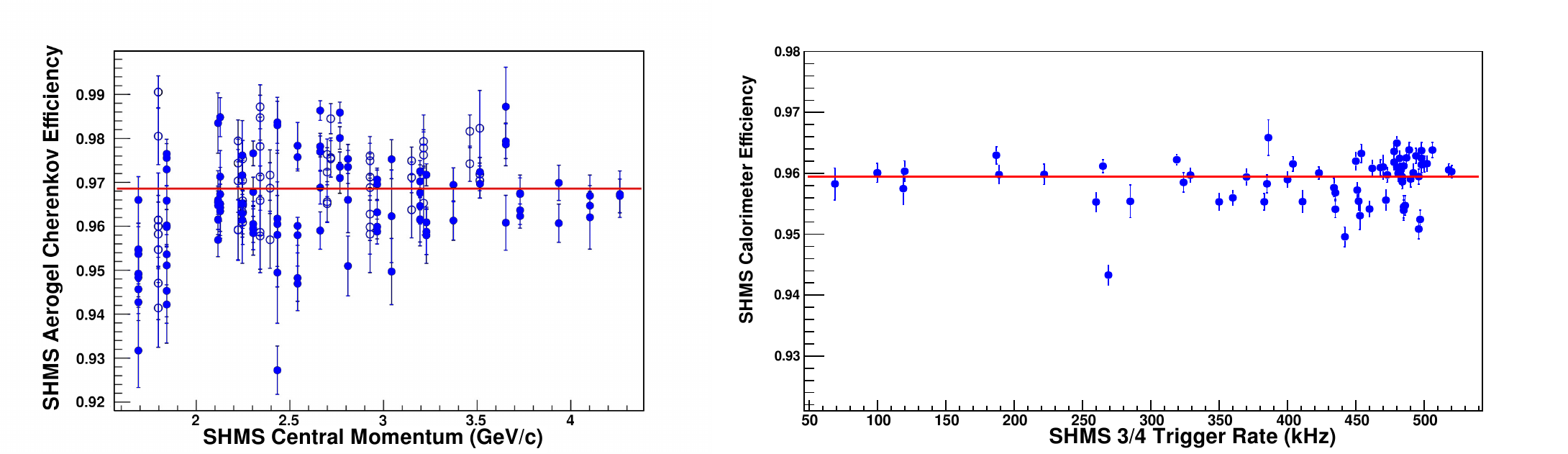}
\caption{ (left) The pion identification efficiency of the SHMS aerogel detector as a function of the pion momentum for $\pi^{+}$ (solid) and $\pi^-$ (open). (right) The SHMS calorimeter efficiency as a function of 3/4 trigger rate ($\pi^{+}$). The solid lines are constant values fits that were used as the efficiency corrections. Only the statistical uncertainties are shown.}
\label{fig:SHMS_aero_eff}
\end{figure*}

\begin{figure*}[htb!]
\includegraphics[width=0.9\textwidth]{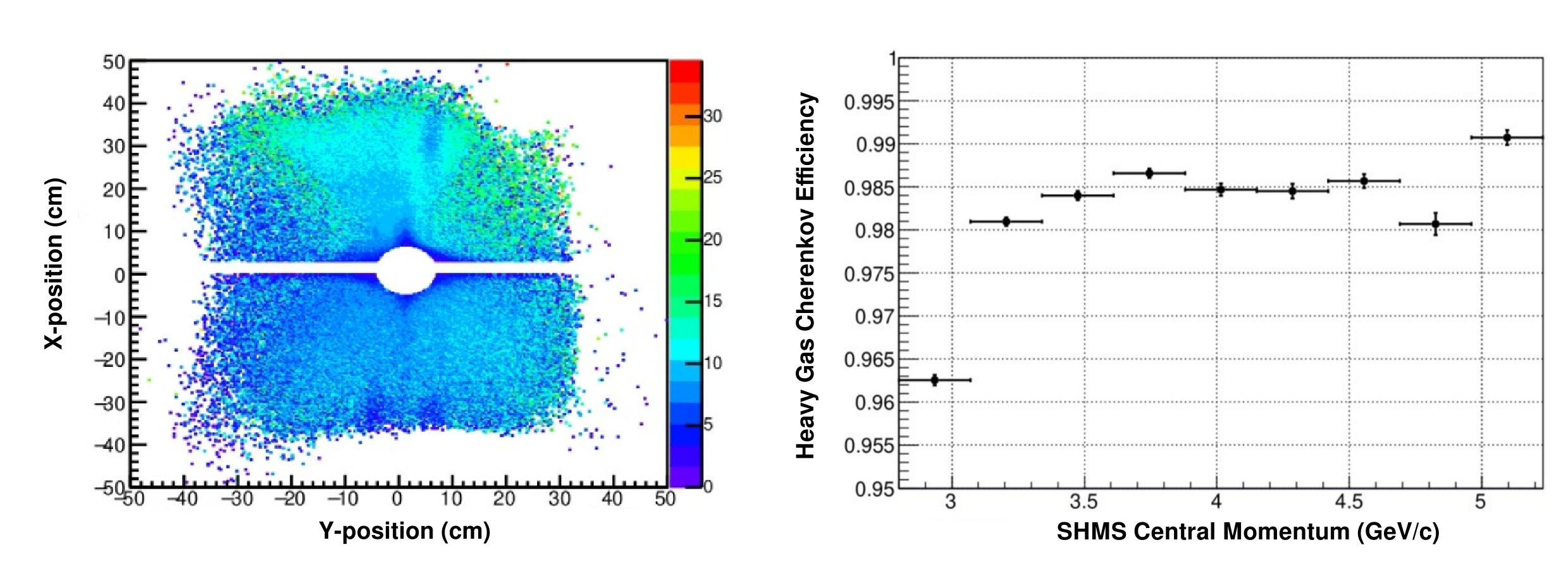} 
    \caption{(left)The x-position vs. y-position of hits on the heavy-gas Cherenkov detector, showing the inefficient region of the Cherenkov detector that was removed from the analysis. The color bar represents the number of photo-electrons. (right) The charged pion efficiency of the heavy-gas Cherenkov detector as a function of the SHMS momentum.}
    \label{fig:hgcer_xvsy}
\end{figure*}

 The pions in the SHMS (the hadron spectrometer) were identified using the electron-hadron
coincidence time, the heavy-gas (C$_4$F$_8$O at less than 1 atm. pressure) 
threshold Cherenkov detector, the aerogel Cherenkov detector~\cite{s3_aero_ref}, and a segmented Pb-glass calorimeter~\cite{s1_cal_ref}.  The pion identification efficiency of the aerogel Cherenkov varied between 94\% for low momentum ($<$~2~GeV/c) pions to 97\% for the highest momentum pions as shown in Fig.~\ref{fig:SHMS_aero_eff} (left). The SHMS calorimeter efficiency was $\sim$ 96\% as shown in Fig.~\ref{fig:SHMS_aero_eff} (right). 
The heavy-gas threshold Cherenkov detector had an inefficient region near the center of the detector. The events from this inefficient region were removed from the analysis using a geometric cut as shown in Fig.~\ref{fig:hgcer_xvsy} (left). The efficiency of the heavy-gas Cherenkov detector above the pion threshold, after removing events from the inefficient region, is shown in Fig.~\ref{fig:hgcer_xvsy} (right). 

\begin{figure*}[htb!]
\includegraphics[width=0.35\textwidth]{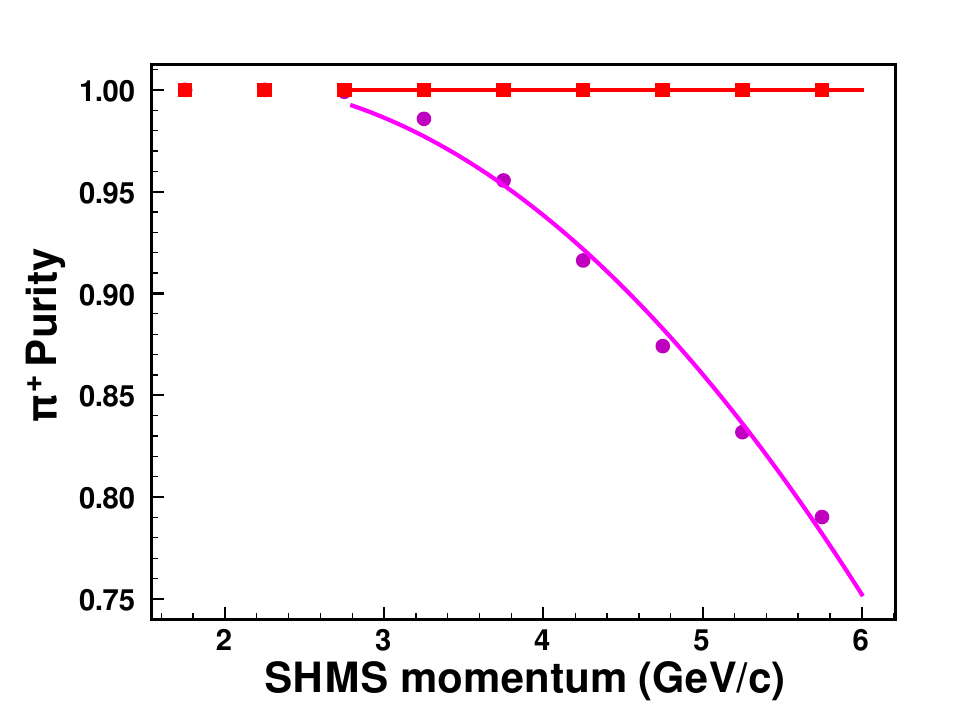}
\includegraphics[width=0.45\textwidth]{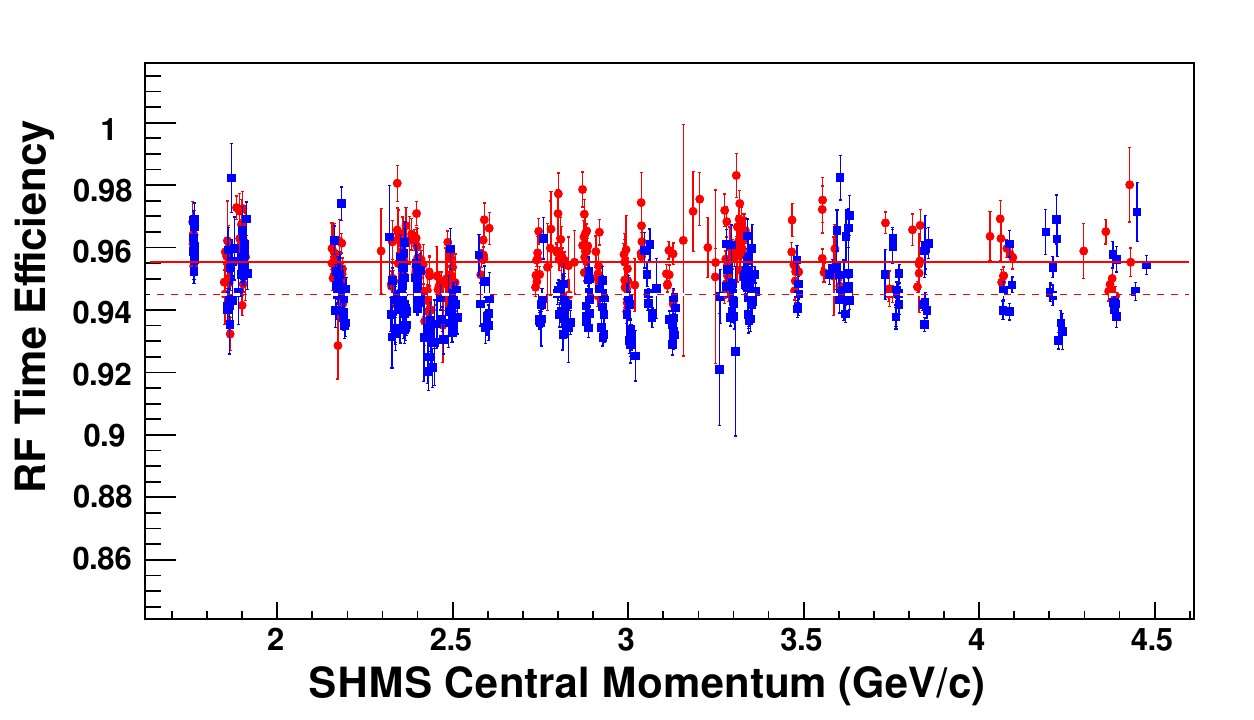}
\caption{ (left) The purity of the pion sample with (red squares) and without (magenta circles) constraints from the heavy-gas Cherenkov as a function of the pion momentum. (right) The RF time efficiency of the $\pi^+$ (blue squares) and $\pi^-$ (red circles) as a function of SHMS central momentum. The lines are the constant value fits for $\pi^+$ (dotted) and $\pi^-$ (solid) with $\chi^2$ per degree of freedom 3.86 and 6.21 respectively. A constant value of 0.95 was used as the RF time efficiency throughout the experiment. Only the statistical uncertainties are shown.}
\label{fig:rf_eff}
\end{figure*}

In addition, the radio-frequency (RF) time information provided for each beam bucket along with electron-hadron coincidence time was also used for particle identification.
The purity of the pion sample was determined using the RF timing information with and without constraints from the heavy-gas Cherenkov, as shown in Fig.~\ref{fig:rf_eff} (left) for the positive pions. Events with positive pion momenta above
2.8 GeV/c have significant kaon contamination when not suppressed by the constraint from the heavy-gas Cherenkov detector. This contamination was negligible for negative pions.
In this analysis, the heavy-gas Cherenkov was used to suppress kaons, therefore, a pion purity of 1.0 was assumed. The difference in the extracted multiplicity, with kaon rejection using the heavy-gas Cherenkov or with a correction to the pion purity when not using the heavy-gas Cherenkov, was used to determine the systematic uncertainty due to kaon contamination of the pion sample. This difference was negligible for negative pions.
The efficiency of the RF constraint as a function of SHMS momentum is shown in Fig.~\ref{fig:rf_eff} (right) for $\pi^{+}$ (black squares) and $\pi^{-}$ (red circles).




The electron-pion 
coincidence events were recorded in approximately 1-hour-long runs via a data 
acquisition system operated using the CEBAF Online Data Acquisition 
(CODA) software package~\cite{s4_coda}. The accidental backgrounds were subtracted by sampling the accidental events corresponding to several adjacent beam buckets on either side of the true coincident events. 
Prescaled singles (inclusive) 
electron and proton events were simultaneously recorded for systematic studies. 

Data collected on the two aluminum foil targets were used to subtract the events from the aluminum walls of the cryogenic target cell.  The background from $\pi^{0}$ production, subsequent decay 
and eventual conversion to electron-positron pairs was determined to be negligible based on representative data collected by detecting positrons in the HMS.
\begin{figure}[hbt!]
\includegraphics[width=0.5\columnwidth]{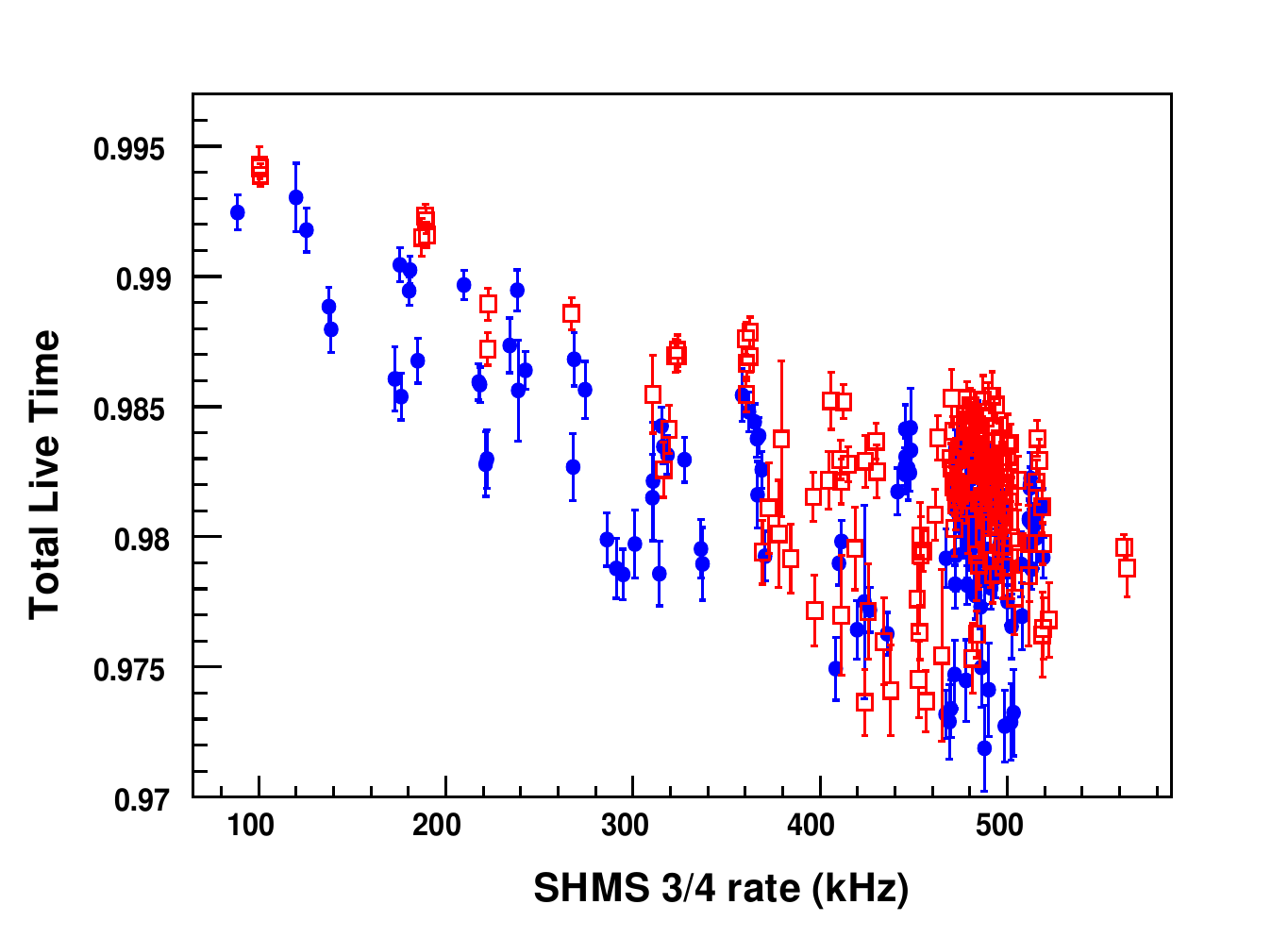}
    \caption{The total live time of the $\pi^{+}$ (red open squares) and $\pi^{-}$ (blue circles) events as a function of the trigger rate in the SHMS which was the hadron spectrometer. Only the statistical uncertainties are shown.}
    \label{fig:live_time}
\end{figure}
The total live-time (product of the electronic and computer live-times) of the data acquisition (DAQ) system was measured using a special trigger called an Electronic Dead Time Monitor (EDTM). The EDTM consists of a known, fixed-frequency trigger, deliberately chosen to be a low rate (10 Hz in this experiment) such that it does not block the real trigger. The ratio of the recorded to the expected EDTM triggers was used as the total live-time of the DAQ. The total live time plotted as a function of the hadron trigger rate in the SHMS spectrometer is shown in~Fig.~\ref{fig:live_time}. 
\section{Data Analysis}
The charge-normalized and background subtracted coincidence yield on the $^1$H and $^2$H targets were obtained by integrating over the experimental phase space, including azimuthal angle $\phi$ and $p_T$. This normalized SIDIS pion electroproduction yield was corrected for the live-time and all the inefficiencies listed earlier and binned in $z$. The corrected yield, along with yields from the Monte Carlo simulation, was used to extract the multiplicity, defined as the ratio of the SIDIS cross section to the DIS cross section for each target (p/d) and charged pion type, given by:
\begin{equation}
M^{\pi^{\pm}}_{\mbox{p/d}}(x,Q^2,z) = \frac{d\sigma_{ee'\pi X}}{d\sigma_{ee'X}}=\frac{\displaystyle\sum_{i} e_{i}^{2} q_{i}^{\mbox{p/d}}(x) D_{q_i \rightarrow \pi^{\pm}}(z)}{ \displaystyle\sum_{i} e_{i}^{2} q_{i}^{\mbox{p/d}}(x)}, 
\label{eq:bprm}
\end{equation}

\begin{figure*}[hbt!]
    \includegraphics[width=0.95\textwidth]{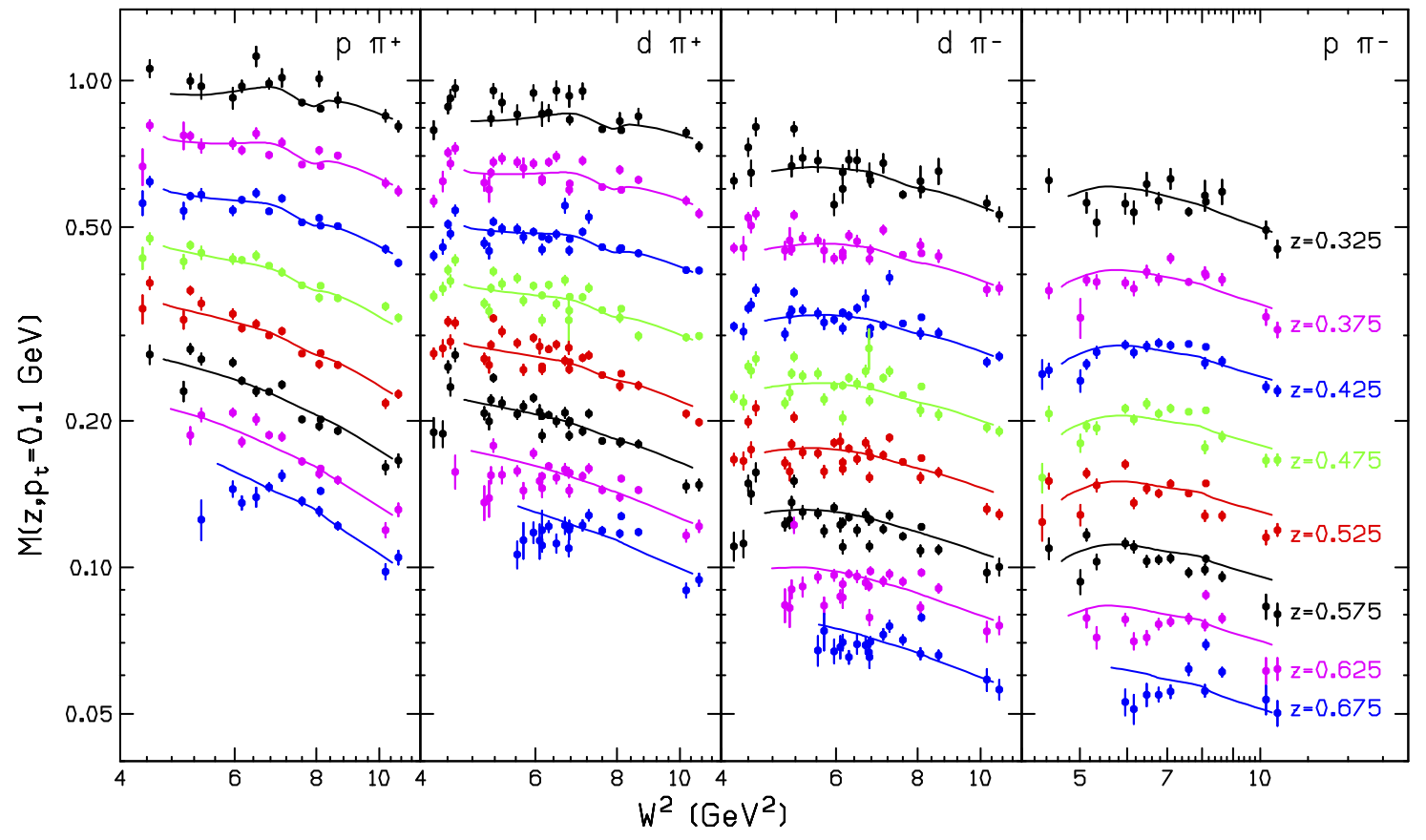}
    \caption{The multiplicities at $p_T=0.1$ GeV, averaged over 
    $\phi^{*}$ as a function of $W^2$ for $z$ bins ranging from $z$ = 0.325 to 0.675. From left to right, the panels are for 
    $\pi^+$ from a proton target, 
    $\pi^+$ from a deuteron target, 
    $\pi^-$ from a deuteron target,  and
    $\pi^-$ from a proton target, 
    The solid lines are from the empirical fits. Only the statistical uncertainties are shown.}
    \label{fig:4mults}
\end{figure*}

The four multiplicities at different values of $z$ are shown as a function of $W^2$ in Fig.~\ref{fig:4mults}.
The four multiplicities show the expected $z$ dependence (i.e decreasing monotonically with increasing $z$). They also show an increase in the slope of the $W^2$ dependence as the $z$ increases.\\

\begin{figure*}[!htb]
    \includegraphics[width=0.9\textwidth]{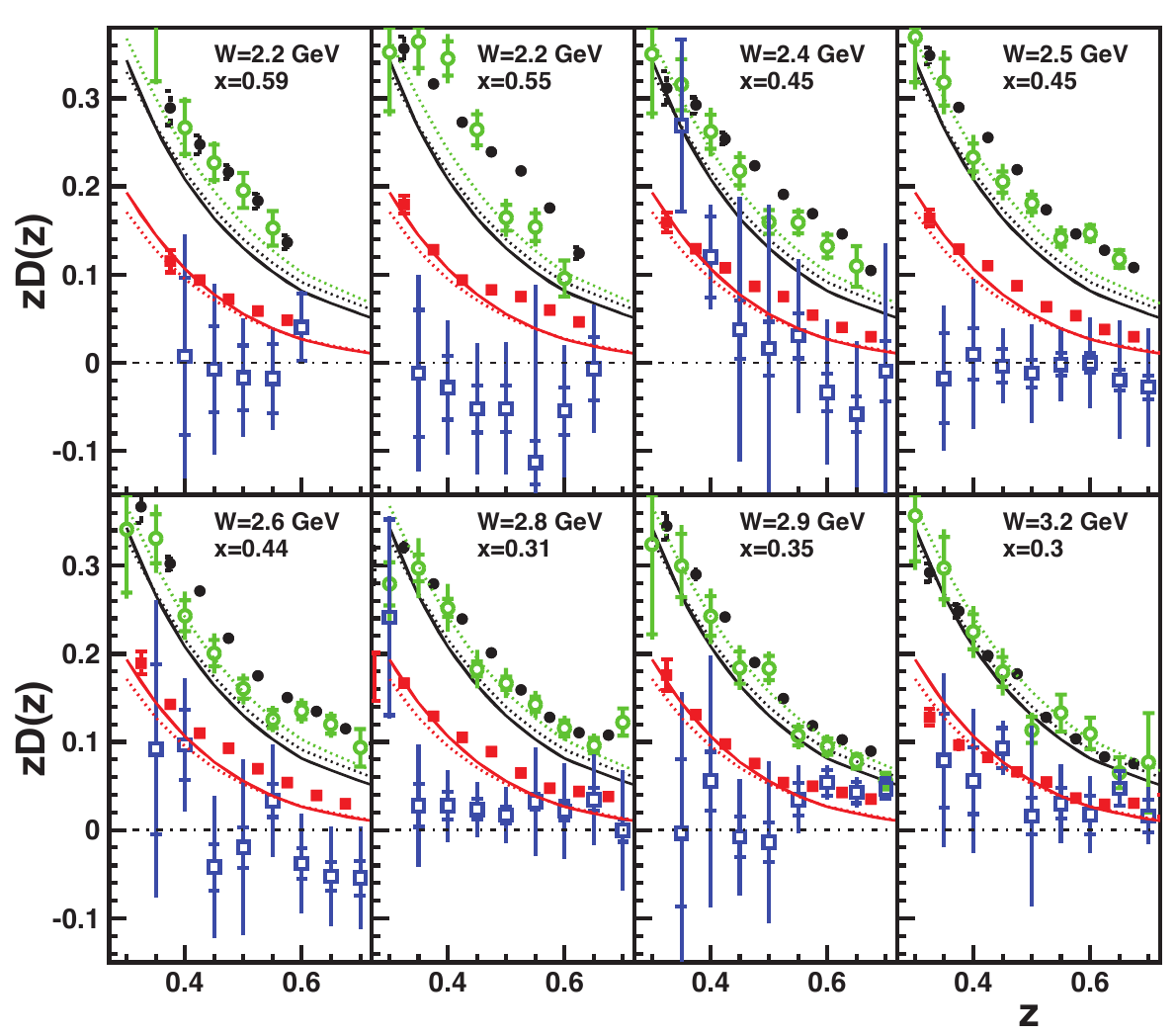}
    \caption{The 4 extracted FF shown as a function of $z$ for the eight kinematic settings. The open (green) and solid (black) circles are the two favored FF, $D_{d \pi^{-}}$ and $D_{u \pi^{+}}$, respectively. While the open (blue) and solid (red) squares are the two unfavored FF, $D_{d \pi^{+}}$ and $D_{u \pi^{-}}$, respectively.  The dashed lines are the results of global fits from DSS~\cite{s6_DSS1,s7_DSS2}, while the solid lines are from the global fit by the JAM~collaboration~\cite{s8_JAM}. Both were calculated for the highest $W$ (3.2 GeV) setting. The JAM collaboration imposes isospin symmetry and hence they produce only one favored FF and one unfavored FF. The inner error bars show the statistical uncertainty while the outer error bars are the total uncertainty which includes the systematic uncertainty in quadrature. The open data points have been shifted relative to the solid points for clarity.
    }
   \label{fig:4ffs}
\end{figure*}

\begin{figure*}[!htb]
    \includegraphics[width=0.9\textwidth]{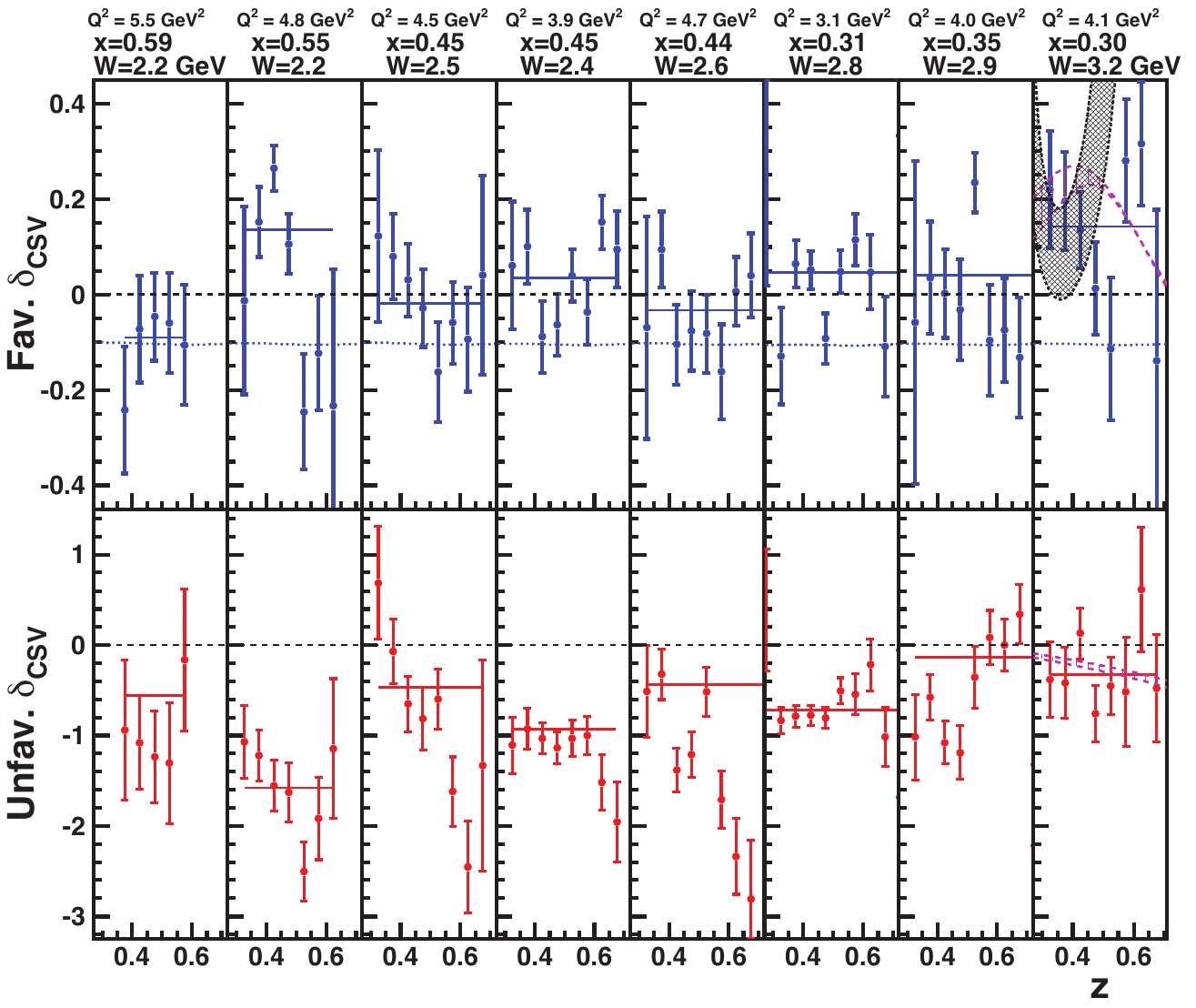}
    \caption{The $z$ dependence of the charge/isospin symmetry violating parameter ($\delta_{CSV}$) for the favored FF (top panels) and un-favored FF (bottom panels), extracted from the measured charged pion multiplicities on hydrogen and deuterium targets. Horizontally, the panels are ordered in decreasing values of $x$ (increasing $W$). All 8 kinematic settings are shown. Assuming charge symmetry, the $\delta_{CSV}$ parameter should be zero, as indicated by the black solid line (top panels). The magenta band with vertical hatching in the last panels is $\delta_{CSV}$ and its uncertainty from the global fit by Peng and Ma~\cite{s9_pandm}, the black band with angled hatching is $\delta_{CSV}$ and its uncertainty from the global fit by the MAP collaboration~\cite{s10_lhapdf},
while the dotted lines are from the DSS global fits~\cite{s6_DSS1,s7_DSS2}. Only statistical uncertainty is shown.}
   \label{fig:ff_asym_all}
\end{figure*}

Assuming charge symmetry for PDF but not for the fragmentation functions (FF), the multiplicity $M^{\pi^{\pm}}_{\mbox{p/d}}(x,Q^2,z)$ can be expanded in terms of the quark content of the two targets as:
\begin{align}
\nonumber
M^{\pi^+}_{\mbox{p}}(x,Q^2,z) =
\frac{4u(x)D_{u \pi^{+}}(z) +\bar{d}(x)D_{d \pi^{-}}(z) } {4u(x)+4\bar{u}(x)+d(x)+\bar{d}(x) + 2s(x)} + 
\frac{d(x)D_{d\pi^{+}}(z) + 4\bar{u}(x)D_{u\pi^{-}}(z) + 2s(x)D_{s\pi^{+}}(z)}{4u(x)+4\bar{u}(x)+d(x)+\bar{d}(x) + 2s(x)} ~~~~~~~~~~~~~~~~~~~~~~~~~~~~~~~~~~~~~~~~~~~~~~~~&&\\ \nonumber
M^{\pi^-}_{\mbox{p}}(x,Q^2,z) =
\frac{ 4\bar{u}(x)D_{u\pi^{+}}(z)+d(x)D_{d \pi^{-}}(z) }
{4u(x)+4\bar{u}(x)+d(x)+\bar{d}(x) + 2s(x)} + 
\frac{\bar{d}(x)D_{d\pi^{+}}(z) + 4u(x)D_{u\pi^{-}}(z) + 2s(x)D_{s\pi^{-}}(z)}{4u(x)+4\bar{u}(x)+d(x)+\bar{d}(x) + 2s(x)} ~~~~~~~~~~~~~~~~~~~~~~~~~~~~~~~~~~~~~~~~~~~~~~~~&&\\ \nonumber
M^{\pi^+}_{\mbox{d}}(x,Q^2,z) =
\frac{[4u(x)+4d(x)]D_{u\pi^{+}}(z)+ [\bar{u}(x)+\bar{d}(x)]D_{d \pi^{-}}(z)}{5[u(x)+\bar{u}(x)+d(x)+\bar{d}(x)] + 4s(x)} +
\frac{[u(x)+d(x)]D_{d\pi^{+}}(z) + 2s(x)D_{s\pi^{+}}(z)}{5[u(x)+\bar{u}(x)+d(x)+\bar{d}(x)] + 4s(x)} + ~~~~~~~~~~~~~~~~~~~~~~~~~~~~~~~~~~~~~~~~~&& \\ \nonumber
\frac{[4\bar{u}(x)+4\bar{d}(x)]D_{u\pi^{-}}(z) + 2s(x)D_{s\pi^{+}}(z)}{5[u(x)+\bar{u}(x)+d(x)+\bar{d}(x)] + 4s(x)} ~~~~~~~~~~~~~~~~~~~~~~~~~~~~~~~~~~~~~~~~~~~~~~~~~~~~~~~~~~~~~~~~~~~~~~~~~~~~~~~~~~~~~~~~~~~~~~~~~~~~~~~~~~~&& \\ \nonumber
M^{\pi^-}_{\mbox{d}}(x,Q^2,z)=
\frac{[4\bar{u}(x)+ 4\bar{d}(x)]D_{u\pi^{+}}(z) + [u(x)+d(x)]D_{d \pi^{-}}(z)}{5[u(x)+d(x)+\bar{u}(x)+\bar{d}(x)] + 4s(x)} + 
\frac{[\bar{u}(x)+\bar{d}(x)]D_{d\pi^{+}}(z)+ 2s(x)D_{s\pi^{-}}(z)}{5[u(x)+d(x)+\bar{u}(x)+\bar{d}(x)] + 4s(x)} + ~~~~~~~~~~~~~~~~~~~~~~~~~~~~~~~~~~~~~~~~~&&\\ \nonumber
\frac{[4u(x)+ 4d(x)]D_{u\pi^{-}}(z) + 2s(x)D_{s\pi^{-}}(z)}{5[u(x)+d(x)+\bar{u}(x)+\bar{d}(x)] + 4s(x)},
~~~~~~~~~~~~~~~~~~~~~~~~~~~~~~~~~~~~~~~~~~~~~~~~~~~~~~~~~~~~~~~~~~~~~~~~~~~~~~~~~~~~~~~~~~~~~~~~~~~~~~~~~~&& \\ 
\refstepcounter{equation}
\addtocounter{equation}{-1}
\label{eq:s1}
\end{align}

where $s(x)=\bar{s}(x)$ are the strange ($s$) quark PDF, $D_{u\pi^{+}}$ and $D_{d\pi^{-}}$ are the favored FF and $D_{d\pi^{+}}$ and $D_{u\pi^{-}}$ are the un-favored FF, respectively, with $u(d)\pi^{\pm}$ representing $u(d) \rightarrow \pi^{\pm}$ and $D_{s\pi^{+}} = D_{s\pi^{-}}$ are the $s$ quark FF. 
Note that, under charge symmetry (CS) these reduce to just one favored and one un-favored FF, since CS implies $D_{u\pi^{+}}=D_{d\pi^{-}}$ and $D_{u\pi^{-}}=D_{d\pi^{+}}$.
The four FF as a function of $z$ are extracted from the four multiplicities by simultaneously solving the system of four equations shown above for the eight kinematic settings listed in Table~\ref{tab:kinematics}. Here we assume that the ratio of longitudinal to transverse cross sections ($R=\sigma_L/\sigma_T$) is flavor independent.
 The CTEQ5~\cite{s5_cteq} PDF were used for $u$ and $d$ quarks while the deFlorian, Sassot, and Stratmann (DSS)~\cite{s6_DSS1,s7_DSS2} PDF and FF were used for the $s$ quark. 
These extracted FF as a function of $z$ are shown in Fig.~\ref{fig:4ffs} for the eight kinematic settings. They are also compared to two different global fits of existing data, one by DSS~\cite{s6_DSS1,s7_DSS2} and the other by the Jefferson Lab Angular Momentum collaboration~(JAM)~\cite{s8_JAM} calculated for the highest $W$ (3.2 GeV) setting.   
Within the experimental uncertainties, the four extracted FF converge to the same values at the lowest $x$ or highest $W$,  over the entire range of $z$ (0.3 - 0.7). 
At the lowest $x$ or highest $W$, they are also in agreement with the global fits. The FF deviate from the global fits as $x$ increases or the $W$ decreases. These results likely point to the importance of higher twist corrections at high $x$ or low $W$ kinematics, which drop off as inverse power laws in $W^2$ and/or $Q^2$.\\
The favored and un-favored  CSV parameter ($\delta_{CSV}$) extracted from the FF are shown in Fig.~\ref{fig:ff_asym_all} for the eight kinematic settings. They are also compared to two different global fits of existing data, one by deFlorian, Sassot, and Stratmann (DSS)~\cite{s6_DSS1,s7_DSS2} and the other by Peng and Ma~\cite{s9_pandm}.
\section{Systematic Uncertainties}

\begin{table}[!hbt]
    \caption{List of systematic uncertainties contributing the uncertainty in the multiplicities.}
    \label{tab:syst}    \centering
    \begin{tabular}{lc}
    \hline
        Source & Uncertainty (\%) \\
        \hline \hline
       Charge & 0.45 \\
        Target density $^{1}$H ($^{2}$H) & 0.7 (0.6) \\
        Target boiling correction & 0.2 \\
        Target end cap subtraction & 0.1 \\
        Tracking efficiency &  0.1 \\
        Live time & 0.04 \\
        Particle identification & 0.8 \\
        Background subtraction & 0.5 \\
        Acceptance  & 0.7 \\
        Kinematics &  0.1 \\
        Radiative correction & 1 \\
        Inclusive cross-section & 2\\
        FADC rate dependence & 0.9  \\ 
        \hline \hline
        Total & 2.8  \\ \hline 
   \end{tabular}
\end{table}


The sources of systematic uncertainties and the total systematic uncertainty of the experiment are listed in Table~\ref{tab:syst}. The systematic uncertainty of the charge measurement was determined from the average variation of the charge between data sets collected under similar experimental conditions. The instrumental uncertainty due to electronic noise in the gain monitoring system was also included. 
There is a 0.7\% (0.6\%) correlated uncertainty due to
uncertainty in the target density for $^1$H ($^2$H), which includes contributions from the uncertainty in the target length, thermal contraction, temperature, pressure, and the equation of state used to calculate the target density. In addition, the uncertainty in the corrections due to local variation in the cryogenic target density was estimated using dedicated scans of the experimental yield with increasing beam current. These scans were carried out before and after the production period of the experiment. The average variation in the current dependence of the measured yield between multiple scans and multiple equivalent analyses along with the residual current dependence of the yield on a carbon foil was used as the systematic uncertainty for the target boiling correction (no current dependent density variation is expected for a carbon foil).
The systematic uncertainty due to tracking efficiency was determined from the average variation of the efficiency between periods with the same trigger rates. The error in the fit parameters of a linear fit of the rate dependence of the live-time correction is used to estimate the systematic uncertainty due to the live-time correction.

The systematic uncertainty in the event selection arising from the particle identification cuts was determined from the average variation in the experimental yield when the cuts were varied by a small fixed amount (typically $\pm$10\% of the nominal values) and between multiple equivalent analyses of the same data set. The systematic uncertainty of the background subtraction procedure arises from the uncertainties in the models used to simulate the various sources of background. This uncertainty was determined from the average variation in the measured yield when the model parameters were varied. The systematic uncertainty due to radiative correction was estimated from the average variation of the correction factor when the generation limits of the simulation of these radiative processes were varied and when the cross section models in the simulation were varied. Additional details on the models of the radiative processes and their uncertainty can be found in Ref.~\cite{s11_Asatur12, s12_bosted_fit}. The systematic uncertainty due to the acceptance model in the Monte Carlo simulation was estimated from the variation of the multiplicity when the acceptance cuts were varied. The uncertainty due to the beam energy, spectrometer momentum, and angle settings (i.e. kinematic) was determined from the average variation of the multiplicities when the kinematic settings were varied by the measurement uncertainty of the beam energy, spectrometer momentum, and angles. The uncertainty in the inclusive cross section is from the latest fits to the world data~\cite{s12_bosted_fit}. The total systematic uncertainty of 2.8\% is the quadrature sum of all uncertainties from the different sources.  \\
\indent
For the sum and difference ratios obtained from the multiplicities, most of these systematic uncertainties cancel to first order and were found to be negligible compared to the statistical uncertainty of the sum difference ratios. Only the correlated uncertainty due to target density and the uncertainty due to the inclusive cross section were the major contributions to the sum and difference ratio and led to a 2.2\% systematic uncertainty for these ratios. The systematic uncertainty of the extracted FF arising from the normalization type systematic uncertainties of the multiplicities was studied by scaling the multiplicities and evaluating the variation in FF. From this study, the systematic uncertainty of FF was determined to be $\sim$4\% for one pair of favored and unfavored FF with small statistical uncertainty. The systematic uncertainty was found to be comparable or smaller than the statistical uncertainty for the pair of FF with larger statistical uncertainty. 
Similarly, the variation in the FF CSV parameters due to the choice of PDF and the normalization uncertainty was used to determine the systematic uncertainties of the $\delta_{\mbox{CSV}}$. They were found to be comparable or smaller than the statistical uncertainties.
\section{Results}
The four multiplicities, $M^{\pi^+}_{p}$, $M^{\pi^+}_{d}$, $M^{\pi^-}_{p}$ and  $M^{\pi^-}_{d}$ obtained using the $\pi^{\pm}$ yield from hydrogen and deuterium targets are listed in 
~~~Table~\ref{tab:mult}. The statistical uncertainty is also listed.


\begin{table*}[!hbt]
\vspace{-7ex}
\caption{The four $p_{T}$ integrated multiplicities at an average $p_T =$ 0.1 GeV. The $\rho^0$ contributions to the pion yields have been subtracted.}
\label{tab:mult}\centering
\begin{tabular}{cccccccccccc}
\hline\hline
$x$ & $Q^2$(GeV$^2$) & $W$ (GeV) & $z$& $M^{\pi^+}_{p}$ & $\delta M^{\pi^+}_{p}$ & $M^{\pi^+}_{d}$ & $\delta M^{\pi^+}_{d}$ &$M^{\pi^-}_{p}$ & $\delta M^{\pi^-}_{p}$ & $M^{\pi^-}_{d}$ & $\delta M^{\pi^-}_{\
d}$ \\
\hline
0.59 &   5.5 &   2.2 &   0.375 &        0.7183 & 0.0369 &       0.6068 & 0.018  &       0.3395 & 0.0239 &       0.4383 & 0.0147\\
0.59 &   5.5 &   2.2 &   0.425 &        0.5373 & 0.0176 &       0.4483 & 0.0089 &       0.2358 & 0.0099 &       0.29   & 0.0066\\
0.59 &   5.5 &   2.2 &   0.475 &        0.4185 & 0.0119 &       0.3448 & 0.006  &       0.1619 & 0.006  &       0.2052 & 0.004 \\
0.59 &   5.5 &   2.2 &   0.525 &        0.3182 & 0.0116 &       0.26   & 0.0056 &       0.1192 & 0.0053 &       0.1523 & 0.0035\\
0.59 &   5.5 &   2.2 &   0.575 &        0.2192 & 0.0099 &       0.1915 & 0.0053 &       0.0845 & 0.0048 &       0.1084 & 0.0032\\
0.55 &   4.8 &   2.2 &   0.325 &        1.0138 & 0.0269 &       0.8554 & 0.023  &       0.5756 & 0.0205 &       0.6476 & 0.0246\\
0.55 &   4.8 &   2.2 &   0.375 &        0.7735 & 0.0116 &       0.6438 & 0.01   &       0.3718 & 0.0078 &       0.4555 & 0.0097\\
0.55 &   4.8 &   2.2 &   0.425 &        0.5826 & 0.0077 &       0.4746 & 0.0065 &       0.2492 & 0.0046 &       0.3272 & 0.0056\\
0.55 &   4.8 &   2.2 &   0.475 &        0.4529 & 0.007  &       0.3659 & 0.0057 &       0.1882 & 0.0038 &       0.2379 & 0.0043\\
0.55 &   4.8 &   2.2 &   0.525 &        0.3648 & 0.0059 &       0.2748 & 0.0047 &       0.1441 & 0.0029 &       0.1647 & 0.0032\\
0.55 &   4.8 &   2.2 &   0.575 &        0.2687 & 0.0059 &       0.2119 & 0.0047 &       0.1044 & 0.0027 &       0.1241 & 0.0032\\
0.55 &   4.8 &   2.2 &   0.625 &        0.1753 & 0.0072 &       0.1445 & 0.0056 &       0.0687 & 0.0034 &       0.0779 & 0.0037\\
0.45 &   3.9 &   2.4 &   0.325 &        0.9321 & 0.0353 &       0.9122 & 0.0263 &       0.5285 & 0.0205 &       0.5999 & 0.021 \\
0.45 &   3.9 &   2.4 &   0.375 &        0.7293 & 0.0146 &       0.6687 & 0.0106 &       0.3741 & 0.0075 &       0.4328 & 0.008 \\
0.45 &   3.9 &   2.4 &   0.425 &        0.5446 & 0.0094 &       0.4777 & 0.0065 &       0.2722 & 0.0044 &       0.3148 & 0.0047\\
0.45 &   3.9 &   2.4 &   0.475 &        0.4241 & 0.0078 &       0.3661 & 0.0052 &       0.1932 & 0.0033 &       0.226  & 0.0035\\
0.45 &   3.9 &   2.4 &   0.525 &        0.3263 & 0.0058 &       0.2863 & 0.004  &       0.146  & 0.0025 &       0.1642 & 0.0026\\
0.45 &   3.9 &   2.4 &   0.575 &        0.2546 & 0.004  &       0.2078 & 0.0034 &       0.099  & 0.002  &       0.1187 & 0.0022\\
0.45 &   3.9 &   2.4 &   0.625 &        0.1958 & 0.0033 &       0.1526 & 0.0032 &       0.0653 & 0.0019 &       0.0798 & 0.0022\\
0.45 &   3.9 &   2.4 &   0.675 &        0.13   & 0.0048 &       0.1025 & 0.0054 &       0.0406 & 0.0032 &       0.0549 & 0.0036\\
0.45 &   4.5 &   2.5 &   0.325 &        0.9717 & 0.0173 &       0.8287 & 0.0148 &       0.5448 & 0.0175 &       0.6215 & 0.0147\\
0.45 &   4.5 &   2.5 &   0.375 &        0.7009 & 0.0085 &       0.6057 & 0.0073 &       0.3733 & 0.0076 &       0.4346 & 0.0068\\
0.45 &   4.5 &   2.5 &   0.425 &        0.5421 & 0.0049 &       0.4636 & 0.0042 &       0.2743 & 0.0039 &       0.3078 & 0.0035\\
0.45 &   4.5 &   2.5 &   0.475 &        0.4123 & 0.0036 &       0.3493 & 0.003  &       0.1957 & 0.0027 &       0.2239 & 0.0024\\
0.45 &   4.5 &   2.5 &   0.525 &        0.2943 & 0.0028 &       0.2503 & 0.0022 &       0.13   & 0.002  &       0.1553 & 0.0018\\
0.45 &   4.5 &   2.5 &   0.575 &        0.2246 & 0.0023 &       0.1905 & 0.0017 &       0.0964 & 0.0017 &       0.1134 & 0.0015\\
0.45 &   4.5 &   2.5 &   0.625 &        0.1747 & 0.0022 &       0.1431 & 0.0015 &       0.064  & 0.0015 &       0.0836 & 0.0014\\
0.45 &   4.5 &   2.5 &   0.675 &        0.1346 & 0.0023 &       0.1068 & 0.0016 &       0.0459 & 0.0016 &       0.0593 & 0.0014\\
0.44 &   4.7 &   2.6 &   0.325 &        1.0511 & 0.0332 &       0.9432 & 0.0273 &       0.614  & 0.0221 &       0.6708 & 0.0225\\
0.44 &   4.7 &   2.6 &   0.375 &        0.7518 & 0.0116 &       0.6804 & 0.0099 &       0.4129 & 0.0072 &       0.4735 & 0.0077\\
0.44 &   4.7 &   2.6 &   0.425 &        0.5708 & 0.007  &       0.4765 & 0.0058 &       0.2773 & 0.004  &       0.314  & 0.0043\\
0.44 &   4.7 &   2.6 &   0.475 &        0.4087 & 0.0053 &       0.3447 & 0.0045 &       0.206  & 0.003  &       0.2319 & 0.0033\\
0.44 &   4.7 &   2.6 &   0.525 &        0.3026 & 0.0039 &       0.266  & 0.0034 &       0.1386 & 0.0021 &       0.1585 & 0.0023\\
0.44 &   4.7 &   2.6 &   0.575 &        0.2263 & 0.0033 &       0.1826 & 0.0028 &       0.0949 & 0.0017 &       0.1086 & 0.0019\\
0.44 &   4.7 &   2.6 &   0.625 &        0.1828 & 0.0029 &       0.1431 & 0.0024 &       0.0669 & 0.0014 &       0.0838 & 0.0016\\
0.44 &   4.7 &   2.6 &   0.675 &        0.1414 & 0.0033 &       0.108  & 0.0026 &       0.0471 & 0.0014 &       0.0613 & 0.0016\\
0.44 &   4.7 &   2.6 &   0.725 &        0.1069 & 0.0065 &       0.093  & 0.005  &       0.0366 & 0.0029 &       0.0469 & 0.003 \\
0.31 &   3.1 &   2.8 &   0.275 &        1.0446 & 0.0363 &       1.0144 & 0.0303 &        0.702 & 0.0449 &       0.7777 & 0.0399\\
0.31 &   3.1 &   2.8 &   0.325 &        0.8734 & 0.0068 &       0.7765 & 0.0055 &       0.5383 & 0.0064 &       0.5742 & 0.0061\\
0.31 &   3.1 &   2.8 &   0.375 &        0.6573 & 0.0039 &       0.5839 & 0.0031 &       0.3791 & 0.0034 &       0.4247 & 0.0033\\
0.31 &   3.1 &   2.8 &   0.425 &        0.4951 & 0.0025 &       0.4393 & 0.002  &       0.2731 & 0.0021 &       0.3077 & 0.002 \\
0.31 &   3.1 &   2.8 &   0.475 &        0.368  & 0.002  &       0.3247 & 0.0015 &       0.1976 & 0.0015 &       0.217  & 0.0014\\
0.31 &   3.1 &   2.8 &   0.525 &        0.2637 & 0.0017 &       0.2354 & 0.0013 &       0.1339 & 0.0012 &       0.1521 & 0.0012\\
0.31 &   3.1 &   2.8 &   0.575 &        0.1897 & 0.0017 &       0.1675 & 0.0014 &       0.0894 & 0.0013 &       0.1049 & 0.0012\\
0.31 &   3.1 &   2.8 &   0.625 &        0.1502 & 0.0018 &       0.1345 & 0.0016 &       0.0714 & 0.0013 &       0.0813 & 0.0011\\
0.31 &   3.1 &   2.8 &   0.675 &        0.1289 & 0.0016 &       0.1095 & 0.0014 &       0.054  & 0.0011 &       0.0614 & 0.001 \\
0.31 &   3.1 &   2.8 &   0.725 &        0.0925 & 0.0025 &       0.0793 & 0.0021 &       0.0365 & 0.002  &       0.0478 & 0.0015\\
0.35 &   4.0 &   2.9 &   0.325 &        0.9467 & 0.0239 &       0.8247 & 0.0221 &       0.578  & 0.0263 &       0.6287 & 0.0292\\
0.35 &   4.0 &   2.9 &   0.375 &        0.7006 & 0.0081 &       0.6287 & 0.0077 &       0.3867 & 0.0079 &       0.4353 & 0.0089\\
0.35 &   4.0 &   2.9 &   0.425 &        0.5027 & 0.0048 &       0.4351 & 0.0049 &       0.2555 & 0.0043 &       0.2919 & 0.0049\\
0.35 &   4.0 &   2.9 &   0.475 &        0.3484 & 0.0038 &       0.2989 & 0.0044 &       0.1726 & 0.0032 &       0.1965 & 0.0036\\
0.35 &   4.0 &   2.9 &   0.525 &        0.2516 & 0.0028 &       0.2256 & 0.0033 &       0.1186 & 0.0021 &       0.144  & 0.0024\\
0.35 &   4.0 &   2.9 &   0.575 &        0.1837 & 0.0022 &       0.1688 & 0.0023 &       0.088  & 0.0016 &       0.0974 & 0.0018\\
0.35 &   4.0 &   2.9 &   0.625 &        0.144  & 0.002  &       0.1311 & 0.0015 &       0.0682 & 0.0012 &       0.0761 & 0.0014\\
0.35 &   4.0 &   2.9 &   0.675 &        0.1158 & 0.0017 &       0.1062 & 0.0013 &       0.0493 & 0.001  &       0.0547 & 0.0012\\
0.35 &   4.0 &   2.9 &   0.725 &        0.0993 & 0.002  &       0.0792 & 0.0012 &       0.0363 & 0.0011 &       0.0382 & 0.0012\\
0.35 &   4.0 &   2.9 &   0.775 &        0.0735 & 0.0039 &       0.0601 & 0.0019 &       0.0235 & 0.0022 &       0.0299 & 0.0021\\
0.30 &   4.1 &   3.2 &   0.325 &        0.8152 & 0.015  &       0.7441 & 0.012  &       0.4599 & 0.0134 &       0.5324 & 0.0128\\
0.30 &   4.1 &   3.2 &   0.375 &        0.5951 & 0.0093 &       0.5383 & 0.0074 &       0.3078 & 0.008  &       0.3606 & 0.0075\\
0.30 &   4.1 &   3.2 &   0.425 &        0.4268 & 0.0049 &       0.3982 & 0.004  &       0.2241 & 0.0041 &       0.2585 & 0.0037\\
0.30 &   4.1 &   3.2 &   0.475 &        0.3258 & 0.004  &       0.2878 & 0.0032 &       0.1574 & 0.0032 &       0.1816 & 0.0028\\
0.30 &   4.1 &   3.2 &   0.525 &        0.2148 & 0.003  &       0.1939 & 0.0024 &       0.1097 & 0.0025 &       0.1209 & 0.0021\\
0.30 &   4.1 &   3.2 &   0.575 &        0.156  & 0.0035 &       0.1389 & 0.0028 &       0.0738 & 0.0031 &       0.0902 & 0.0029\\
0.30 &   4.1 &   3.2 &   0.625 &        0.1182 & 0.0029 &       0.1103 & 0.0023 &       0.0545 & 0.0024 &       0.0659 & 0.0023\\
0.30 &   4.1 &   3.2 &   0.675 &        0.0947 & 0.0025 &       0.0843 & 0.002  &       0.0447 & 0.0023 &       0.0481 & 0.002 \\
0.30 &   4.1 &   3.2 &   0.725 &        0.0809 & 0.0042 &       0.0656 & 0.0029 &       0.0365 & 0.0082 &       0.0453 & 0.0045\\\hline \hline
\end{tabular}
\end{table*}